\documentclass[10pt]{article}

\usepackage{amsmath}
\usepackage{array}
\usepackage{appendix}
\usepackage{graphicx}
\usepackage{amsfonts}
\usepackage{amssymb}
\usepackage{mathrsfs}
\usepackage{yfonts}
\usepackage{euscript}
\usepackage{upgreek}
\usepackage{slantsc}
\usepackage{calligra}
\usepackage[T1]{fontenc}
\usepackage{epsf}
\usepackage{latexsym}

\usepackage{tipa}

\def\be{\begin{equation}}
\def\ee{\end{equation}}
\def\beq{\begin{equation}}
\def\eeq{\end{equation}}
\def\bea{\begin{eqnarray}}
\def\eea{\end{eqnarray}}

\def\ni{\noindent}
\def\foo{\footnote}

\def\hat{\widehat}
\def\tilde{\widetilde}
\def\!{\hspace{-1.6667em}}

\def\mD{\mbox{D}}

\def\mK{\mbox{K}}

\def\mM{\mbox{M}}
\def\mN{\mbox{N}}

\def\mR{\mbox{R}}

\def\mX{\mbox{X}}

\def\me{\mbox{e}}

\def\mg{\mbox{g}}
\def\mh{\mbox{h}}

\def\mm{\mbox{m}}

\def\mp{\mbox{p}}

\def\bupSigma{\mbox{\boldmath$\Sigma$}}                 
\def\sbSigma{\mbox{\scriptsize\boldmath$\Sigma$}}

\def\fA{\mbox{\sffamily A}}
\def\fB{\mbox{\sffamily B}}

\def\ux{\underline{{x}}}

\def\bh{\underline{\underline{\mbox{h}}}  }            

\def\bh{\mbox{{\bf h}}}

\def\scC{\mbox{\scriptsize ${\cal C}$}}          
\def\scE{\mbox{\scriptsize ${\cal E}$}}          
\def\scF{\mbox{\scriptsize ${\cal F}$}}
\def\scH{\mbox{\scriptsize ${\cal H}$}}          
\def\scI{\mbox{\scriptsize ${\cal I}$}}

\def\scL{\mbox{\scriptsize ${\cal L}$}}          
\def\scM{\mbox{\scriptsize ${\cal M}$}}          
\def\scN{\mbox{\scriptsize ${\cal N}$}}
\def\scO{\mbox{\scriptsize ${\cal O}$}}

\def\scR{\mbox{\scriptsize ${\cal R}$}}          
\def\scS{\mbox{\scriptsize ${\cal S}$}}

\def\scU{\mbox{\scriptsize ${\cal U}$}}          

\def\iB{\mbox{\scriptsize$B$}}   

\def\FrQ{\mbox{\Large $\mathfrak{q}$}}
\def\FrT{\mathfrak{T}}                            

\def\FrP{\mbox{\Large\boldmath$\mathfrak{p}$}}    
\def\FrH{\mbox{\boldmath$\mathfrak{H}$}}          

\def\FrT{\mbox{\boldmath$\mathfrak{T}$}}                        

\def\sFrG{\mbox{\boldmath\scriptsize$\mathfrak{g}$}}

\def\FrM{\mbox{\Large $\mathfrak{m}$}}                         
\def\FrMgen{\mbox{\boldmath$\mathfrak{M}$}}                     

\def\tFrT{\mbox{\normalsize $\mathfrak{T}$}}

\def\FrR{\mbox{\Large $\mathfrak{r}$}}

\def\FrS{\mbox{\Large $\mathfrak{s}$}}

\def\FrG{\mbox{\Large $\mathfrak{g}$}}                            
 
\def\sa{\mbox{\scriptsize a}}
\def\sb{\mbox{\scriptsize b}}
\def\scc{\mbox{\scriptsize c}}
\def\sd{\mbox{\scriptsize d}}
\def\se{\mbox{\scriptsize e}}

\def\sg{\mbox{\scriptsize g}} 
\def\sh{\mbox{\scriptsize h}} 
\def\si{\mbox{\scriptsize i}}

\def\sll{\mbox{\scriptsize l}}  
\def\sm{\mbox{\scriptsize m}}
\def\sn{\mbox{\scriptsize n}} 
\def\so{\mbox{\scriptsize o}} 
\def\sp{\mbox{\scriptsize p}}

\def\sr{\mbox{\scriptsize r}}
\def\sss{\mbox{\scriptsize s}}  
\def\st{\mbox{\scriptsize t}}

\def\sG{\mbox{\scriptsize G}}
\def\sH{\mbox{\scriptsize H}}

\def\sS{\mbox{\scriptsize S}}

\def\sX{\mbox{\scriptsize X}}

\def\sfA{\mbox{\sffamily{\scriptsize A}}}      
\def\sfB{\mbox{\sffamily{\scriptsize B}}}      
\def\sfC{\mbox{\sffamily{\scriptsize C}}}      
\def\sfF{\mbox{\sffamily{\scriptsize F}}}      
\def\sfG{\mbox{\sffamily{\scriptsize G}}}      
\def\sfL{\mbox{\sffamily{\scriptsize L}}}      
\def\sfS{\mbox{\sffamily{\scriptsize S}}}      
\def\sfT{\mbox{\sffamily{\scriptsize T}}}      

\def\to{\mbox{\tiny o}}
\def\tp{\mbox{\tiny p}}

\def\K{Kucha\v{r} }

\def\NSI{Na\"{\i}ve Schr\"{o}dinger Interpretation }
\def\CPI{Conditional Probabilities Interpretation }

\def\pa{\partial}
\def\d{\textrm{d}}

\def\5Star{\mbox{\Large$\star$}}              
\def\sumi2{\sum\mbox{}_{\mbox{}_{\mbox{\scriptsize $i$=1}}}^2}
\def\sumi3{\sum\mbox{}_{\mbox{}_{\mbox{\scriptsize $i$=1}}}^3}

\def\sumj3{\sum\mbox{}_{\mbox{}_{\mbox{\scriptsize $j$=1}}}^3}
\def\sumk3{\sum\mbox{}_{\mbox{}_{\mbox{\scriptsize $k$=1}}}^3}

\begin{document}

\begin{titlepage}

\begin{center}

{\Huge{\bf SPACES OF SPACES}}

\vspace{.1in}

{\bf Edward Anderson} 

\vspace{.1in}

{\em DAMTP, Centre for Mathematical Sciences, Wilberforce Road, Cambridge CB3 OWA.} \normalsize

\end{center}

\begin{abstract}

Wheeler emphasized the study of Superspace($\bupSigma$) -- the space of 3-geometries on a spatial manifold of fixed topology $\bupSigma$.  
This is a configuration space for GR; knowledge of configuration spaces is useful as regards dynamics and QM. 
In this Article I consider furthmore generalized configuration spaces to all levels within the conventional 
`equipped sets' paradigm of mathematical structure used in fundamental Theoretical Physics.
This covers A) the more familiar issue of topology change in the sense of topological manifolds (tied to cobordisms), including via pinched manifolds.
B) The less familiar issue of not regarding as fixed the yet deeper levels of structure: topological spaces themselves (and their metric space subcase), collections of subsets and sets.
Isham has previously presented quantization schemes for a number of these.

In this Article, I consider some classical preliminaries for this program, aside from the most obvious (classical dynamics for each).
Rather, I provide I) to all levels Relational and Background Independence criteria, which have Problem of Time facets as consequences.
I demonstrate that many of these issues descend all the way down, whilst also documenting at which level the others cease to apply. 
II) I consider probability theory on configuration spaces. 
In fact such a stochastic treatment is how to further mathematize the hitherto fairly formal and sketchy subject of records theory (a type of formultion of quantum gravity). 
Along these lines I provide a number of further examples of records theories.
%
%
To this example I now add 1) \v{C}ech cohomology, 2) Kendall's random sets, 3) the lattice of topologies on a fixed set.
I finally consider 4) sheaves, both as a generalization of \v{C}ech cohomology and in connection to the study of stratified manifolds such as Superspace($\bupSigma$) itself.  

\end{abstract}

\end{titlepage}

\section{Introduction} 

\subsection{Spaces of spaces}

Wheeler \cite{Battelle} emphasized the study of Superspace($\bupSigma$) -- the space of 3-geometries on a spatial manifold of fixed topology $\bupSigma$.  
This is a type of configuration space for GR.

\mbox{ } 

\ni {\it Configuration space} $\FrQ$ \cite{Lanczos} is the space of all possible configurations $Q^{\sfA}$ of a classical system.  
As a first simpler example, the most obvious configuration space for a single point particle coincides with its position in space; both are $\mathbb{R}^d$ in dimension $d$. 
However, for $N$ particles, 
the position of each in space is described by $\mathbb{R}^d$ but the most obvious configuration space $\FrQ(N, d)$ now takes the distinct form $\mathbb{R}^{Nd}$ \cite{Lanczos}.  
Moreover, the preceding representation contains absolute space information which can be reduced out \cite{Kendall, FileR}. 
Thus one arrives at reduced configuration spaces. 
Most straightforwardly, passing to the centre of mass frame by quotienting out the translations $Tr(d)$, the relative configuration space $\FrR(N, d) = \mathbb{R}^{nd}$ for $n = N - 1$.
If scale is regarded as absolute feature so that the dilations $Dil$ are to be quotiented out too, one passes to {\it preshape space} $\FrP(N, d) = \mathbb{S}^{nd - 1}$.
If rotations $Rot(d)$ are quotiented out as well, one passes to {\it shape space}  $\FrS(N, d)$; 
this is just $\mathbb{S}^{n - 1}$ in 1-$d$ or $\mathbb{CP}^{n - 1}$ in 2-$d$ but not known to be systematically evaluable for all $N$ in 3-$d$.  
If one re-introduces scale at this point, one passes to {\it relational space} ${\cal R}(N, d)$ which is the cone over the preceding (and furthermore just $\mathbb{R}^n$ in 1-$d$). 
Finally for the particular case of the triangle of particles in 2-$d$, $\mathbb{CP}^1 = \mathbb{S}^2$, the cone over which is $\mathbb{R}^3$ albeit not with flat metric \cite{FileR}.

{            \begin{figure}[ht]
\centering
\includegraphics[width=0.75\textwidth]{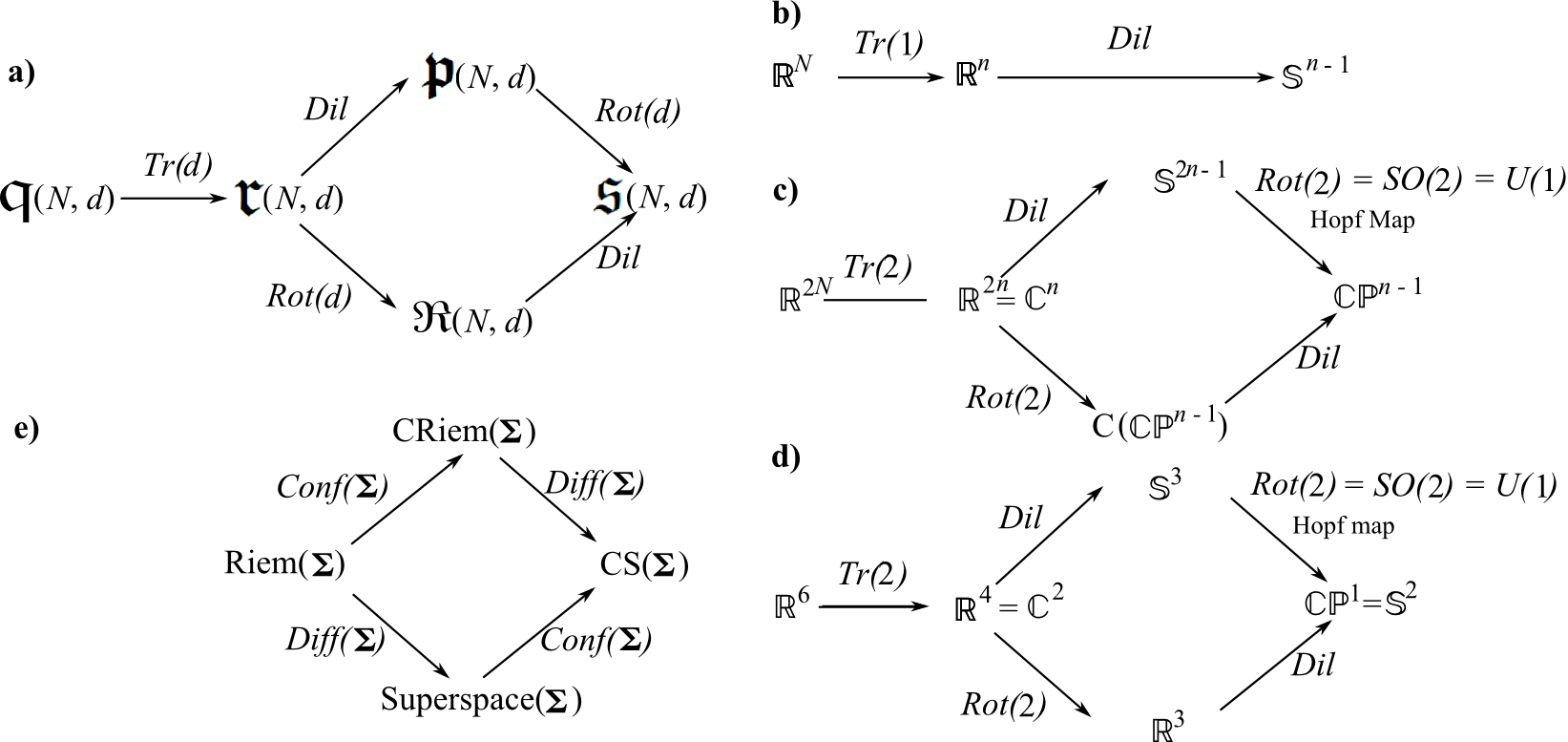}
\caption[Text der im Bilderverzeichnis auftaucht]{        \footnotesize{a) Relational particle mechanics configuration spaces.
b) The particular mathematical spaces these take in 1-$d$, which simplify due to there being no nontrivial continuous rotations here.
c) The particular mathematical spaces these take in 2-$d$; C denotes `cone over'.
d) Further specialization in the case of the triangle.  
e) Configuration spaces for GR.} }
\label{Std-Config-Spaces}\end{figure}    }

\ni Now returning to the GR case \cite{DeWitt}, the simplest albeit redundant configuration space is Riem($\bupSigma$): the space of spatial 3-metrics on $\Sigma$. 
Then if one quotients out $Diff$($\bupSigma$) the diffeomorphisms of $\bupSigma$, one arrives at 
\beq
\mbox{Superspace($\bupSigma$) = Riem($\bupSigma$)/$Diff$($\bupSigma$) }.
\label{Superspace}
\eeq
\ni If the conformal transformations $Conf$($\bupSigma$) on $\bupSigma$ are quotiented out instead, one arrives at CRiem($\bupSigma$).
If both of these are quotiented out, one arrives at {\it conformal superspace} CS($\bupSigma$) \cite{York74, FM96}; 
CRiem($\bupSigma$) is then termed {\it pointwise conformal superspace} \cite{FM96}.  

\mbox{ } 

\ni Wheeler's primary motivation for studying Superspace($\bupSigma$) comes from QM unfolding on configuration space, by which other cases of configuration space are interesting too.
[This insight is missed by those only studying single-particle models.]
Furthermore, as we shall see below, Wheeler argued that the competing notion of spacetime dissolves at the quantum gravitational level, 
due to which he considered adhering to dynamical first principles instead. 

\mbox{ } 

\ni Moreover, studying specifically Superspace($\bupSigma$) involves specifically the metric and differential geometry level of structure. 
On the other hand, CRiem($\bupSigma$) is at the level of conformal metric structure, and CS($\bupSigma$) is at the level of conformal metric and differential structure.  
Recollect at this point the levels of mathematical structure usually assumed in Theoretical Physics [Fig 2.a); see Sec 4 for an outline of their specific features].
This is the `equipped sets' paradigm of mathematics (various alternatives to which are outlined in Secs \ref{Commentary} and \ref{Last}.  
In the spatial metric version of this, Superspace($\bupSigma$) is then one of the corresponding spaces of spaces, 
but there are clearly many others such, and not just the often also evoked ones of Fig 1.
Fig 2.b) lists these and Sec 3 outlines each in turn (alongside further variants).  
Such an idea of generalized configuration spaces has been advocated by Isham \cite{I84b, I89-Rev, I89-Latt, IKR, I91, I03}.
In this Article I point out that this amounts to a substantial generalization of Wheeler's question 
-- into the cornucopia of questions in Secs 6-11 with furthermore there being a lack of good reasons to go no further down the set of levels than Superspace($\bupSigma$).
Indeed, this descending of the levels may be viewed as recursive application of Wheeler's `looking for zeroth principles to explain first principles' \cite{Battelle}.

{            \begin{figure}[ht]
\centering
\includegraphics[width=1.0\textwidth]{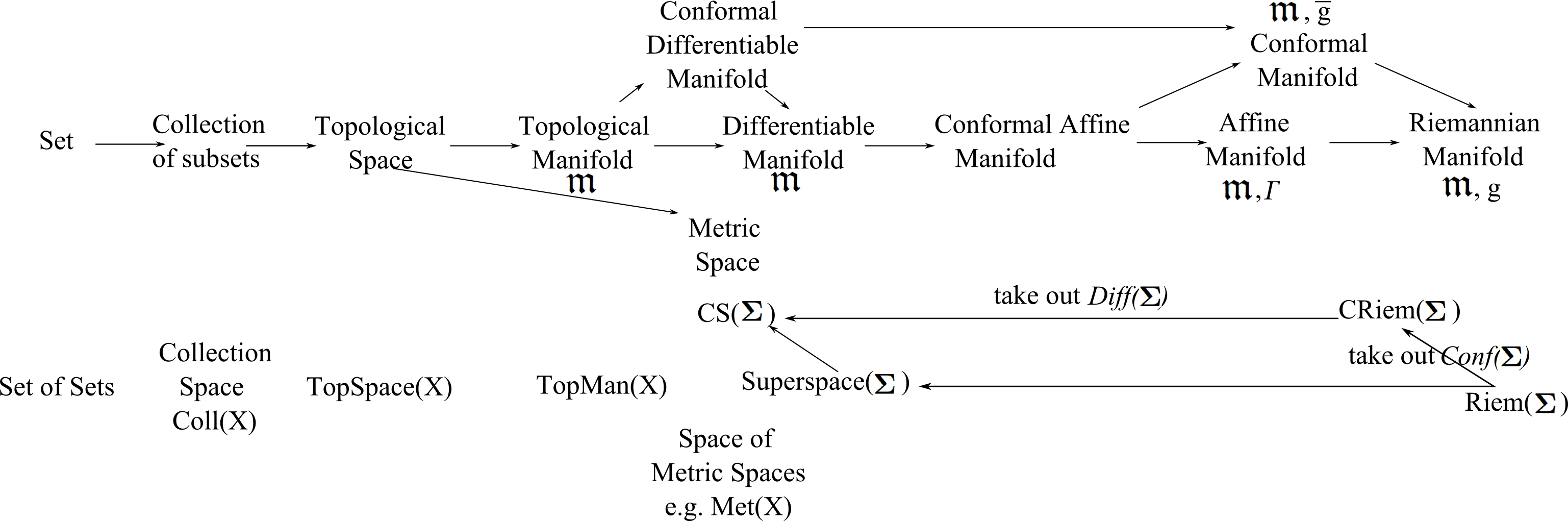}
\caption[Text der im Bilderverzeichnis auftaucht]{        \footnotesize{a) Levels of mathematical structure commonly assumed in Classical Physics.  
Secs 3 and 4 consider allowing for each of these to be dynamical in turn.  
Note that `metric structure = conformal structure + localized scale', and then in the indefinite spacetime metric case, conformal structure additionally amounts to causal structure.   
b) The spatial case version's corresponding spaces of spaces (omitting the spaces of affine structures until Sec 3.3).} }
\label{Bigger-Set-2}\end{figure}            }
%
\ni Isham \cite{I89-Latt, IKR, I91, I03, I10, ToposRev} considered {\it generalized configuration spaces} in the context of his program of 
{\sl quantizing each level in turn, for each level of mathematical structure in Theoretical Physics}.
Since such a program carries major Background Independence connotations, 
Sec 2 prepares by outlining Background Independence and the consequent Problem of Time at the usual metric and differentiable structure level.
Note also that passing from levels of structure to the corresponding spaces of spaces does not in general keep one within the original set of structures. 
E.g. the space of topologies on a fixed set is a lattice.
Neither does reduction, because quotienting in general does not preserve mathematical structures.
E.g. reduction in general sends manifolds to stratified manifolds. 
See Sec 5 for more. 
{\sl The current Article principally concerns classical precursors for Isham's program}.

\mbox{ }

\ni Precursor I) is the most obvious such (and {\sl not} covered in the current Article): working out what classical dynamics precedes such quantization schemes.
For instance, on the one hand for relational mechanics $N$ particles in 1-$d$ and in 2-$d$ are widely solvable \cite{FileR}.
On the other hand, in 3-$d$ solvability an $N$ by $N$ basis which rapidly becomes complicated and intractable even for small $N$.
To some extent this difference is rooted on the first hand's much simpler and systematically evaluable configuration space topology and metric geometry.

\mbox{ } 

\ni Precursor II) Another classical issue, moreover, concerns how to formulate probability theory on configuration spaces.
In the case of $N$ points in 2-$d$ this is covered by shape statistics -- based on a type of {\it Stochastic Geometry}.  
I identified as a concrete mathematical realization of a {\it Records Theory} -- 
a hitherto rather sketchy conceptual framework from the Quantum Gravity literature, which I explain in Sec 3.
This rather motivates studying probability theory on a wider range of configuration spaces in Secs 8 and 9. 
These include use of Kendall's random sets \cite{Kendall74}, 
study of the lattice of topologies on a fixed set \cite{Lattice, I89-Latt}, \v{C}ech cohomology \cite{Cech} and sheaves \cite{Sheaves, Iversen86, BanaglBook}.  

\mbox{ } 

\ni Precursor III) The classical-level study of {\it relational} and {\it background-independent} features at each level of mathematical structure, which I cover in Secs 6 and 7.    
These features have as consequence the notorious {\it Problem of Time} \cite{Kuchar92, I93}, which thus becomes formulated and addressed to all levels of mathematical structure!
Though as per the article's title, I concentrate on the space of spaces rather than the space of spacetimes, 
or on the two-way map between these, leaving space for future papers on these topics.

\mbox{ }

\ni This Introduction's set of questions amount to my enlarging Wheeler's question along lines provided by the works of Kendall \cite{Kendall74, Kendall}, Isham, 
Kucha\v{r} \cite{Kuchar92} and Barbour \cite{B94I}.  
In investigating II) and III) in the current article, I also touch upon categories, topoi and, especially, sheaves and stratified manifolds \cite{Whitney65, Fischer70, BanaglBook}.  
I end with a brief return to the quantum level in Sec 10; the subsequent Conclusion lists not only the present Article's main advances but also some further questions for the future.
	
\vspace{10in}		

\section{Outline of Background Independence and the Problems of Time}

Problems of Time (PoT) are ubiquitous in background-independent approaches to Quantum Gravity.  
There are multiple such problems \cite{Kuchar92, I93, APoT, APoT2, APoT3}.  
Moreover almost all of them are already present \cite{BI, APoT3} in classical background-independent \cite{A6467Giu06} precursors.
Moreover, many of these problems are less severe in classical precursors. 
I view as an advantage since this area of study is largely unresolved, 
so model arenas are appropriate \cite{Kuchar92, I93, KieferBook, APoT, APoT2, FileR, APoT3} and these include classical background independent description. 
I then found that Barbour-type Relationalism \cite{BB82, B94I, RWR} at the classical level addresses the classical precursors of two of the nine traditional facets of the Problem of Time.
[This is clear in the first two background independent aspects considered below.]
I subsequently constructed an extension of that approach to the other classically-realized facets \cite{BI, FileR, A13, ABook}.

\mbox{ }

\ni See \cite{APoT3} for uncoupled aspects and facets. See \cite{AGates} for an outline of the present approach's couplings, 
and \cite{Kuchar92, I93, APoT} for more extensive (but not complete) treatment of other approach's couplings

\mbox{ } 

\ni The below outline includes the original usually quantum-level PoT facet names for guidance as regards earlier literature \cite{Kuchar92, I93}.  
This is at the conventionally-presented metric to differential geometry level.

\mbox{ } 

\ni 1) {\bf Temporal Relationalism} is that there is no time at the primary level for the universe as a whole.
This can be implemented at the level of action principles as follows.

\ni i) These are not to contain any extraneous times or extraneous time-like variables (Newtonian time is an example of the first and ADM \cite{ADM} lapse of the second).

\ni ii) Time is not to be smuggled into the action in the guise of a label either.

\mbox{ } 

\ni As a specific implementation, make use of manifestly reparametrization invariant actions such as the Jacobi action for Mechanics \cite{Lanczos}. 
GR can indeed also be formulated in this manner \cite{RWR}, 
and furthermore in parametrization-irrelevant form or the geometrical action form  that is dual to that and which does not even make any mention of parameters \cite{AM13, TRiPoD}.
This reparametrization invariance gives rise to GR's quadratic Hamiltonian constraint 

\ni \beq
\scH := \mN_{ijkl}\mp^{ij}\mp^{kl} - \sqrt{\mh}\{  \mR(\ux; \bh] - 2\Lambda\} = 0 \mbox{ } .  
\eeq 
[By an argument of Dirac, reparametrization invariance implies at least one primary constraint].
In the Relational Approach, one is to view this -- and the corresponding Mechanics expression that is usually viewed as an energy equation -- as {\it equations of time}, 
which I denote \scC\scH\scR\scO\scN\scO\scS.
[I use small calligraphic font to pick out constraints.]
Thus the implementation is  

\ni \beq
\mbox{ Absense of a meaningful parameter time } \Rightarrow \scC\scH\scR\scO\scN\scO\scS \mbox{ } .  
\eeq
The {\it emergent Machian time resolution} of Temporal Relationalism is that {\sl time is to be abstracted from change}.  
Change is $\d Q^{\sfA}$, playing a role that replaces that of velocities with respect to external or physically irrelevant label times \cite{TRiPoD}.  
The subsequent Machian emergent time is of the form

\ni\beq
f = F[Q^{\sfA}, \d Q^{\sfA}] \mbox{ } .  
\label{t-em-Mach}
\eeq
A more specific proposal for this involves a sufficient totality of locally relevant change (STLRC).  
Such an expression can indeed be obtained by rearranging $\scC\scH\scR\scO\scN\scO\scS$ into the form (\ref{t-em-Mach}).

\mbox{ } 

\ni $\scH$ in turn leads to the quantum-level {\it Wheeler--DeWitt equation} \cite{DeWitt, Battelle}

\ni\be
\widehat{\scH}\Psi = 0 \mbox{ } 
\ee
This is a subcase of time-independent Schr\"{o}dinger equation (TISE) $\widehat{H}\Psi = E\Psi$, i.e. a stationary alias frozen equation.  
Moreover, this occurs in a situation in which one might expect a time-dependent Schr\"{o}dinger equation $\widehat{H}\Psi = i\hbar\,\pa\Psi/\pa t$ for some notion of time $t$.
Thus this exhibits a {\it Frozen Formalism Problem}, now viewed as not particularly surprising since the classical precursor was already frozen.
Moreover, the classical resolution above does not hold, but e.g. a new semiclassical expression along the lines of \ref{t-em-Mach} can also be constructed.  

\mbox{ }

\ni 2) {\bf Configurational Relationalism} is that 
i) one is to include no extraneous configurational structures either (spatial or internal-spatial metric geometry variables that are fixed-background rather than dynamical).
 
\ni ii) Physics in general involves not only a $\FrQ$ but also a $\FrG$ of transformations acting upon $\FrQ$ that are taken to be physically redundant.
This is a matter of practical convenience: often $\FrQ$ with redundancies is simpler to envisage and calculate with.
For ii), if $\FrG$ is internal, this is another formulation of conventional Gauge Theory. 
The spatial case is similar, in that it can also be thought of as a type of Gauge Theory for space itself.

\mbox{ } 

\ni A general indirect implementation of this involves replacing objects $O$ by 

\ni \beq
\mbox{\Large S}_{g \in \sFrG} \mbox{Maps} \,\, \circ \stackrel{\rightarrow}{{\FrG}_g} O \mbox{ } : 
\label{G-Act-G-All}
\eeq
a $\FrG$-{\it act} $\FrG$-{\it all} construct, in which the group action's (denoted by an overhead arrow) involvement of a $\FrG$ auxiliary variable $g$ 
                   is subsequently cancelled out by a move $\mbox{\Large S}_{g \in \sFrG}$ which runs over the entirety of $\FrG$. 
Examples of such a move include averaging (as in group averaging), taking a sup or inf, or the below case of extremization. 
This method is free to include whatever maps in between the two halves of the procedure.

\mbox{ } 

\ni {\it Best Matching} \cite{BB82} is implementing Configurational Relationalism at the level of Lagrangian variables ($\mbox{\boldmath $Q$}$, $\dot{\mbox{\boldmath $Q$}}$). 
This involves pairing $\FrQ$ with a $\FrG$ such that $\FrQ$ is a space of redundantly-modelled configurations.
Here $\FrG$ acts on $\FrQ$ as a {\it shuffling group}: 
one considers pairs of configurations, keeping one fixed and shuffling the other (i.e. an active transformation) until the two are brought into maximum congruence.
In more detail, one proceeds via constructing a $\FrG$-corrected action.
For the examples considered here, this involves replacing each occurrence of $\dot{\mbox{\boldmath $Q$}}$ with 
$\dot{\mbox{\boldmath $Q$}} - \stackrel{\rightarrow}{\FrG_g}\mbox{\boldmath $Q$}$, where $\stackrel{\rightarrow}{\FrG_g}$ indicates group action.
Then varying with respect to the $\FrG$ auxiliary variables $g^{\sfG}$ produces the {\it shuffle constraints} $\scS\scH\scU\scF\scF\scL\scE_{\sfG}$. 
These arise as {\sl secondary constraints}: via use of equations of motion. 
Being linear in the momenta, their form is denoted by $\scL\scI\scN_{\sfG}$ (a subcase of such linear constraints since not all $\scL\scI\scN_{\sfL}$ arise from shuffling).
Then varying with respect to the $\FrG$ auxiliary variables $g^{\sfG}$ produces the {\it shuffle constraints} $\scS\scH\scU\scF\scF\scL\scE_{\sfG}$. 
These are linear in the momenta.
In the Best Matching procedure, one furthermore takes $\scS\scH\scU\scF\scF\scL\scE_{\sfG}$ as equations in the Lagrangian variables to solve for the $g^{\sfG}$ auxiliaries themselves.
One then substitutes the extremizing solution back into the original action to obtain a reduced action on the 

\ni\beq
\tilde{\FrQ} = \FrQ/\FrG
\eeq
configuration space.  
The summary form of this implementation is thus 

\ni \beq
\mbox{Best Matching} \rightarrow \scS\scH\scU\scF\scF\scL\scE_g = 0 \mbox{ } .  
\eeq

\ni The geometrodynamical subcase of Best Matching, for which $\FrG$ is the spatial diffeomorphisms Diff($\bupSigma$) with corresponding linear constraint the GR momentum constraint 

\ni\be
\scM_i := - 2\mD_j {\mp^j}_i = 0
\ee
is known as the {\it Thin Sandwich}.
This name is best explained by first considering the {\it thick sandwich}.
This prescribes knowns $\mh_{ij}^{(1)}$ and $\mh_{ij}^{(2)}$ on two hypersurfaces -- the `slices of bread' -- from which one is to solve for the finite region of `filling' 
in between (Fig 3.b), in analogy with the QM set-up of transition amplitudes between states at two different times \cite{WheelerGRT}.
However, this turns out to be very ill-defined mathematically.
Due to this, one passes to its `thin limit' \cite{WheelerGRT}, 
with spatial metric $\mh_{ij}$ and its label-time velocity $\dot{\mh}_{ij}$ prescribed as data on a spatial hypersurface $\bupSigma$ (Fig 3.c).
Here the {\it thin sandwich equation} -- $\scM_i$ re-expressed in Lagrangian variables $\mh_{ij}$, $\dot{\mh}_{ij}$, 
including taking an emergent position \cite{BSW} on the form of the lapse -- is taken as an equation to be solved for $\upbeta^i$. 
From this, one constructs an infinitesimal piece of spacetime to the future of $\Sigma$ via forming the extrinsic curvature combination.  
The Thin Sandwich remains a problem -- one of the many facets of the PoT \cite{Kuchar92, I93} -- because the above mathematics remains hard \cite{TSC12}.  

\mbox{ } 

\ni If not resolved at the classical level, supplement $\widehat{\scC\scH\scR\scO\scN\scO\scS}\Psi = 0$ by $\widehat{\scS\scH\scU\scF\scF\scL\scE}_{g}\Psi = 0$ 
(in particular $\widehat{\scH}\Psi = 0$ by $\widehat{\scM}_i\Psi = 0$).

\mbox{ } 

\ni 3) {\bf Constraint Closure} concerns whether the above ${\scH}$ and ${\scM}_i$ (or more generally $\scC\scH\scR\scO\scN\scO\scS$ and $\scS\scH\scU\scF\scF\scL\scE_{g}$)
are all the constraints, or whether these lead to further constraints.
This is assessed by a Dirac-type procedure \cite{Dirac, HTBook, TRiPoD}. 

\ni\beq
\{{\scC}_C, {\scC}_{C^{\prime}}\} \mbox{ } \mbox{ closes } .  
\label{CC}
\eeq
At the quantum level for field theories (in terms of operators and commutator brackets) 
this takes the form of a {\it Functional Evolution Problem}, which is the traditional PoT facet name for this.  
Here,

\ni \beq
\widehat{\scC}_{\sfC} \Psi = 0 \mbox{ } \not{\Rightarrow \mbox{}} \mbox{ } \mbox{\bf [}\widehat{\scC}_{\sfC}\mbox{\bf ,} \, \widehat{\scC}_{\sfC^{\prime}}\mbox{\bf ]}\Psi = 0 
\mbox{ }.
\label{Fun-Evol}
\eeq
automatically as well, so more work may be required. 
Instead, more constraint terms might be unveiled, or the right-hand-side of the second of these equations might be an anomaly term rather than zero.  
For GR in general, this remains an unsolved problem.  

\mbox{ } 

\ni 4) {\bf Expression in terms of Beables}.  
Having found constraints and introduced a classical brackets structure, one can then ask which objects have zero classical brackets with `the constraints'.
These objects, termed {\it observables} or {\it beables} \cite{Bell, ABeables}, are more physically useful than just any $Q^{\sfA}$ and $P_{\sfA}$
due to containing physical information only.  
Applied instead to one constraint and two observables or beables determines that the observables or beables themselves form a closed algebraic structure.  
In this sense, the observables or beables form an algebraic structure that is associated with that formed by the constraints themselves.  
That they are in general hard to find constitutes the {\it Problem of Beables} (a previous name for which is the {\it Problem of Observables}).           
This concerns finding enough quantities $\iB_{\sfB}$ to describe the physics, these {\it beables} being defined as commutants with all of a theory's first-class constraints 

\ni 
\beq
\mbox{\bf \{} \scC_{\sfF}\mbox{\bf ,} \, \iB_{\sfB}\mbox{\bf \}} `=' 0
\label{D-B}
\eeq 
-- Dirac beables -- or maybe just with the linear ones -- \K beables

\ni \beq
\mbox{\bf \{}\scF\scL\scI\scN_{\sfF}\mbox{\bf ,} \, \iB_{\sfB}\mbox{\bf \}} `=' 0
\label{K-B}
\eeq
[These are the first-class linear constraints, which are usually the same as the shuffle constraints, in a functioning theory in the Relational Approach.] 


\ni At the quantum level, the corresponding criteria are built from operator versions and commutator brackets.  
For GR in general, this problem is, once again, an unresolved one \cite{ABeables}.  

\mbox{ } 

\ni 5) {\bf Spacetime Relationalism} 
a.i) There are to be no extraneous spacetime structures, in particular no indefinite background spacetime metrics. 
Fixed background spacetime metrics are also more well-known than fixed background space metrics. 

\ni a. ii) Now as well as considering a spacetime manifold $\FrM$, consider also a $\FrG_{\sS}$ of transformations acting upon $\FrM$ that are taken to be physically redundant.
For GR,  $\FrG_{\sS}$ = $Diff$($\FrM$).

\ni b)  $\FrG_{\sS}$'s generators are to close.

\ni c) Spacetime observables are then quantities forming zero brackets with the $Diff$($\FrM$) generators.  

\mbox{ }

\ni {\it Histories alternative}.
Here one considers finite paths instead of instantaneous changes.
Moreover, histories carry further connotations than paths: for now at classical level, they possess their own conjugate momenta and brackets.
These are Isham--Linden \cite{IL} style histories; the older Gell-Mann--Hartle \cite{GMH, Hartle} style histories are purely quantum.
Histories are another approach with more quantum level than classical motivation.
N.B. that histories have a mixture of spacetime properties and canonical properties.  

\mbox{ } 

\ni 6) {\bf Foliation Independence} is a part of coordinate independence.
Classical GR succeeds to have this desirable property by {\it Refoliation Invariance}: 
that evolving via the dashed or the dotted surface in Fig \ref{Facet-Intro-5}.e) gives the same answer \cite{T73} [Fig 3.f affirms this for classical GR].
The {\it Foliation Dependence Problem} concerns whether this remains the case in further theories, 
in particular at the quantum level for which no counterpart of the above resolution is yet known.   

\mbox{ } 

\ni 7) {\bf Spacetime Construction}. This refers to classical spacetime emerging from assumptions of just space and/or a discrete ontology: Fig \ref{Facet-Intro-5}.f).
It is formerly know as reconstruction due to the ingrainedness of presupposing the existence of spacetime structure.  
\cite{AM13} details its classical level form, whereas it is further motivated at the quantum level, as per Fig \ref{Facet-Intro-5}.g).  

\mbox{ } 

\ni 8) {\bf Global Validity}.           
In \K and Isham's reviews \cite{Kuchar92, I93}, this consists of difficulties with choosing an `everywhere-valid' timefunction (see \cite{ABook} for an update); 
this could refer to being defined over all of space or over all of the notion of time itself.  
They term this the {\it Global Problem of Time}. 
I extend it to mean global difficulties with any of the facets involved, for which I use the plural form {\it Global Problems of Time}.  

\mbox{ } 

\ni 9) {\bf Reconcileability of Multiplicites}.
The absense of this is termed {\it Multiple Choice Problems} \cite{Kuchar92, I93}. 
These are only relevant once the quantum level is under consideration, and, as Fig \ref{Facet-Intro-5}.i) illustrates, 
consists of canonical equivalence of classical formulations of a theory not implying unitary equivalence of the quantizations of each \cite{Gotay00}.  
By this, different choices of timefunction can lead to inequivalent quantum theories.
%
{            \begin{figure}[ht]
\centering
\includegraphics[width=0.85\textwidth]{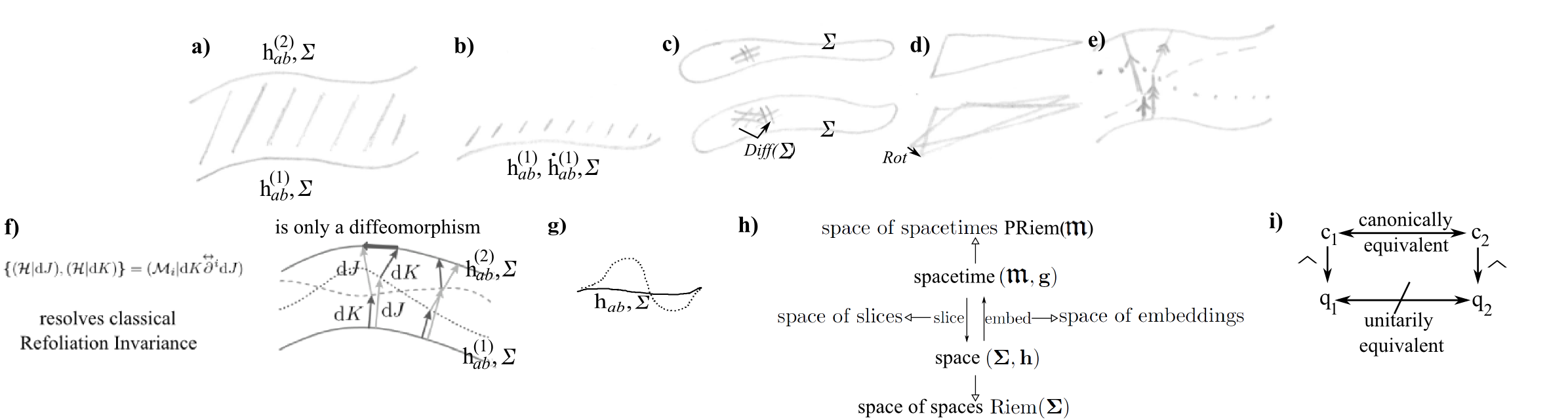}
\caption[Text der im Bilderverzeichnis auftaucht]{        \footnotesize{ a) ADM split of the spacetime metric.
b) Thick Sandwich and its Thin Sandwich limit c).  
The data are as given and the problems to solve are for the spacetime in each shaded region. 
d) is the Thin Sandwich's reworking as the geometrodynamical case of Best Matching: with respect to the spatial diffeomorphisms Diff($\Sigma$).  
e) depicts the geometry of the text's statement of the Foliation Dependence Problem.
f) depicts the resolution of the Foliation Dependence Problem in classical GR: the pictorial form of the bracket of two $\scH$'s closes up the diagram with a mere spatial diffeomorphism. 
g) depicts the dynamical object -- the spatial 3-geometry (solid) -- and the subsequent quantum fluctuations of this, (dotted) which do not all fit into the one spacetime.
%
%
This view of the world is entirely accepted to break down in quantum physics due to Heisenberg's Uncertainly Principle; 
in QM, worldlines are replaced by the more diffuse notion of wavepackets. 
Wheeler then pointed out \cite{Battelle, W79} that in GR, the uncertainty principle now applies to the quantum operator counterparts of $\mh_{ij}$ and $\mp^{ij}$. 
But by the momentum--extrinsic curvature relation of geometrodynamics, this means that $\mh_{ij}$ and the extrinsic curvature $\mK_{ij}$ are not precisely known.   
h) The eightfold cross. 
One passes from spacetime to space by considering a slice and projecting spacetime entities onto it, or by foliating the spacetime with a collection of spaces.
Moving in the opposite direction (harder due to assumption of less structure) involves embedding rather than projecting, and is a construction of spacetime.
It is then indeed valuable to consider not only the space of spaces but also the space of spacetimes, the space of slicings and the space of embeddings.  
Finally, i) supports the statement of the Multiple Choice Problem. 
Here `c' stands for classical formulation, `q' for quantum formulation and $\widehat{\mbox{ }}$ denotes quantization map.} }
\label{Facet-Intro-5} \end{figure}          }

\ni The first seven form `a local theory of background independent', at the metric and differentiable structure level. 
In Sec 6, I consider which of these aspects of Relationalism and Background Independence can be addressed in level-independent manner. 
Then in Sec 7, I consider which features cease to be meaningful or true at some level.

\section{Strategies, in particular Timeless Records and the role of Stochastic Mathematics therein}

Strategies for resolving the above-mentioned frozen facet of the Problem of Time include \cite{Kuchar92, I93, APoT2} the following. 

\mbox{ } 

\ni Strategy A) Perhaps GR has {\bf emergent} rather than fundamental time. 
As a particular case, following Mach `time is to be abstracted from change'. 
Furthemore, all changes should have a chance to contribute (Sec 2's STLRC) \cite{APoT3}. 

\mbox{ } 

\ni Strategy B) Perhaps instead one could see how much of Physics can be envisaged {\it timelessly}.
This {\it solipsist position} involves supplanting `becoming' with `being' at the primary level \cite{Reichenbach, Page1Page2, B94IIEOT, Records}.   
In this sense, Timeless Approaches consider space or the instant as primary and change, dynamics, history and spacetime as secondary.
The \NSI \cite{NSI} then considers simple questions of being, and the \CPI \cite{PW83} conditioned propostions (concerning one type of being conditioned upon another type).
Then conditioning by `the clock reads this time' can be reformulated purely in terms of correlation \cite{PW83}, 
and becoming can be replaced by correlations involving memories within a system's last relevant instant \cite{Page1Page2}.

\mbox{ } 

\ni {\bf Records Theory} \cite{PW83, B94IIEOT, GMH, H99, Records, AKendall} involves localized subconfigurations of a single instant of time.  
It concerns issues such as whether these contain useable information, are correlated to each other, and whether a semblance of dynamics or history arises from this.  
This requires \cite{Records, AKendall}

\ni i) {\it notions of localization} in space and in configuration space \cite{FileR}. 

\ni ii) {\it Notions of information} and {\it correlation}.  
Whilst the notion of distance concept can be extended to the configuration space geometry, 
I also noted {\it patterns} outside of what conventional physicists' notions of information and of correlation encompassed. 
I now identify these with {\it Stochastic Mathematics}.   
Here one is to allot probabilities to timeless propositions.

\mbox{ } 

\ni Example: Relational mechanics in 2-$d$.  
Here, I found \cite{FileR, AKendall} that the key question of `what is the geometry of the underlying configuration space' had already been addressed by Kendall \cite{Kendall}, 
despite his not studying mechanics or dynamics at all.
Instead, he arrived at his shape geometry instead by pursuing probability theory on configuration spaces (what I term Precursor II). 
Furthermore, his geometric probability theory on shape space, and subsequent shape statistics in effect addressed the issue of what Records Theory is from a mathematical perspective.  
Whilst `only a classical treatment', I noted \cite{AKendall} that this went further in other ways than existing quantum gravitationally motivated works on records.  
In particular, he considered $N$-point constellations in 2-$d$ probed by triples: geometrical probability on the shape sphere \cite{Kendall}.  
Indeed, geometrical probability has only recently been brought to attention in the Theoretical Physics literature; 
this has been formulated both on manifolds and on stratified manifolds \cite{Kendall}. 

\mbox{ }

\ni Note the difference between just any geometry and shape geometry: the latter applies to a particular space of spaces. 
There is consequently a difference between just any geometrical probability and statistics versus shape probability and statistics. 
This distinction extends to the probability and statistics on whatever level of mathematical structure's general configuration space alias space of spaces being significant: 
{\it space of spaces stochastics and statistics (SSSS)}. 
The current Article identifies and brings together what is known about this more general class of problems, structures and methods.

\mbox{ }

\ni Strategy C) Perhaps histories are primary instead: {\bf Histories Theory}. 
Now one is to allot probabilities to propositions formulated in terms of histories.

\mbox{ } 

\ni I consider Strategy A) foremost.  
A semiclassical consideration suffices as regards the major practical goal of whether the small inhomogeneities of observational cosmology \cite{HallHaw} 
(galaxies and microwave background hot-spots) are quantum cosmological imprints as amplified by inflation.
In this `slightly inhomogenous cosmology' regime, slow, heavy $h$ modes (scalefactor and homogeneous matter modes) provide an approximate time by which the other fast, light $l$ 
modes (small anisotropies and small inhomogeneities or lumps) evolve.

\mbox{ }

\ni However, strategy A) turns out to have need of Strategies B) and C) for support at the quantum level.  
Namely, that histories decohereing (self-measuring) provides the semiclassical regime. 
Moreover, also 1) histories contain timeless records \cite{GMH, H99}: in this setting, records are but a special 1-time case of histories. 
2) Both records and histories involve coarse and fine graining operations.  
Amongst theoretical physicists, this is better known for histories \cite{Hartle}, though I also established this to be the case for records \cite{Records, FileR}.
Such graining operations are yet furtherly much more well known to occur in Statistical Mechanics.  
The widespread occurrence in mathematics of the graining concept is organized in Sec 8; 
adopting this case, the occurrence of all the other cases is clear from a structural perspective.
3) Histories, but not records by themselves, are tied to decoherence issues \cite{Hartle}.
The elusive question of which degrees of freedom decohere which others in Quantum Cosmology can be 
addressed via where information is actually stored, i.e. where the records are.
These interprotections have prompted me to extend the work of Halliwell \cite{H03H09} 
on all of A) and C) as a {\bf combined strategy} to the configurationally relational case with Machianly interpreted emergent time \cite{AHall, A13}.

\section{Brief outline of levels of mathematical structure}

Here is a more detailed account of the structures of Fig 2 (or Fig \ref{Bigger-Set-2c}, now with the corresponding morphism between them).

\mbox{ }

\ni For the purposes of this Article, by a {\it set} $\mX$ we just mean a collection of elements.  

\mbox{ } 

\ni In the standard paradigm of mathematics, a set $\mX$ can then be equipped with extra layers of structure $\sigma$; I denote this by $\langle \mX, \sigma\rangle$.
Furthemore, one can equip an already pre-established equipped space $\FrS$ (rather than just a set) with extra layers of structure $\langle $\FrS$, \sigma\rangle$. 
Adding equipment is what I mean by levels of mathematical structure, within the traditional if in some ways restrictive context that the bottom layer consists of set theory.
See below for examples, and Secs \ref{Commentary} and \ref{Last} for alternative paradigms.  

\mbox{ }

\ni A {\it topological space} $\langle \mX, \tau\rangle$ is a set $\mX$ \cite{Lee} is a set $\mX$ equipped with a collection of its open subsets, $\tau$, which has the following properties 

\mbox{ } 

\ni Top-1) The union of any collection of these subsets is also in $\tau$.

\ni Top-2) The intersection of any finite number of these subsets is also in $\tau$.  

\ni Top-3) $\mX, \emptyset \in \tau$.

\mbox{ } 

\ni Whereas the above is a very useful and often used type of collection of subsets, N.B. that this is but one of a number of different possible collections of subsets.
{\it Covers} are another example. 
A collection of open sets $\{{\cal O}_{\sfA}\}$ is an {\it open cover} for $\mX$ if $\mX = \bigcup_{\sfA} {\cal O}_{\sfA}$. 
Then a subcollection of an open cover that is still an open cover is a {\it subcover}, $\{{\cal O}_{\sfB}\}$ for $\fB$ a subset of the indexing set $\fA$. 
These are important e.g. through entering the definition of the analytically useful notion of {\it compactness}: that every cover has a finite subcover.
Yet further examples are filters, the $\sigma$-algebras \cite{AMP} used in Measure Theory, and trapping systems as defined in Sec \ref{T}
Thus collections of subsets are a more general level of structure than specifically topological spaces (Fig 4).  
From a mathematical perspective, topological spaces are as far as much of Analysis can be generalized.

\mbox{ } 

\ni A well-known specialization of topological spaces is to metric spaces.
\ni A {\it metric space} $\langle \mX, \mbox{Dist}\rangle$ is a set $\mX$ 
equipped with a {\it metric function} Dist$: \mX \times \mX \rightarrow \mathbb{R}$ satisfying the following properties.

\mbox{ } 

\ni Met-1) $\mbox{Dist}(x, y) \geq 0 \mbox{ } \forall \mbox{ } x, y \in \mX \mbox{  (non-negativity) } .$  

\ni Met-2) $\mbox{If } \mbox{Dist}(x, y) = 0, \mbox{ then } x = y  \mbox{ (separation) } .$

\ni Met-3 $\mbox{Dist}(x, y) = \mbox{Dist}(y, x) \mbox{ (symmetry) } .$

\ni Met-4) $\mbox{Dist}(x, y) \leq \mbox{Dist}(x, z) + \mbox{Dist}(z, y) \mbox{ (triangle inequality) } .$

\mbox{ } 

\ni This structure permits yet further parts of standard analysis to apply.  

\mbox{ } 

\ni Another widely useful specialization is to {\it topological manifolds}: a topological space that has the following properties.  

\mbox{ }

\ni Man-1) {\it Hausdorff}:        for $x, y \in \mX, x \neq y$, $\exists$ open sets ${\cal O}_x, {\cal O}_y \in \tau$ 
                                   such that $x \in {\cal O}_x, y \in {\cal O}_y$ and ${\cal O}_x \bigcap {\cal O}_y = \emptyset$. 

\ni Man-2) {\it Second-countable}: there is a countable collection of open sets such that every open set can be expressed as a union of sets in this collection. 

\mbox{ } 

\ni Note that these axioms perform a selection of the middle ground.
I.e. Hausdorffness guarantees not too small and second-countability guarantees not too large. 
Though this is not necessarily the only such balance point to take.
Thus one might consider a somewhat wider range of topological manifolds \cite{Counter-Top} with middling properties by considering slightly different countability and separation axioms. 
(Hausdorffness is an example of the latter.)

\mbox{ } 

\ni Man-3) {\it Locally Euclidean}: every point $x \in \mX$ has a neighbourhood ${\cal N}_x$ that is homeomorphic to the Euclidean space $\mathbb{R}^p$ (meaning that we can use charts).  

\mbox{ }

\ni Topological manifolds are an extension of the notion of continuity.

\mbox{ }

\ni On the other hand extending the notion of differentiability as well specializes these to differentiable manifolds. 
Here the chart concept is used to everywhere locally harness $\mathbb{R}^n \rightarrow \mathbb{R}^n$ function calculus.  
This amounts to further equipping topological manifolds with differentiable structure.

\mbox{ }

\ni One can further equip differentiable manifolds with affine structure, conformal structure and metric structure in the sequence indicated in Fig \ref{Bigger-Set-2c}.
One usually arrives at conformal structure by stripping down a more well known level of structure to its conformal equivalence classes.
This last step being well-known in Theoretical Physics we simply provide references \cite{Wald, Stewart, AMP}.
%
{            \begin{figure}[ht]
\centering
\includegraphics[width=0.8\textwidth]{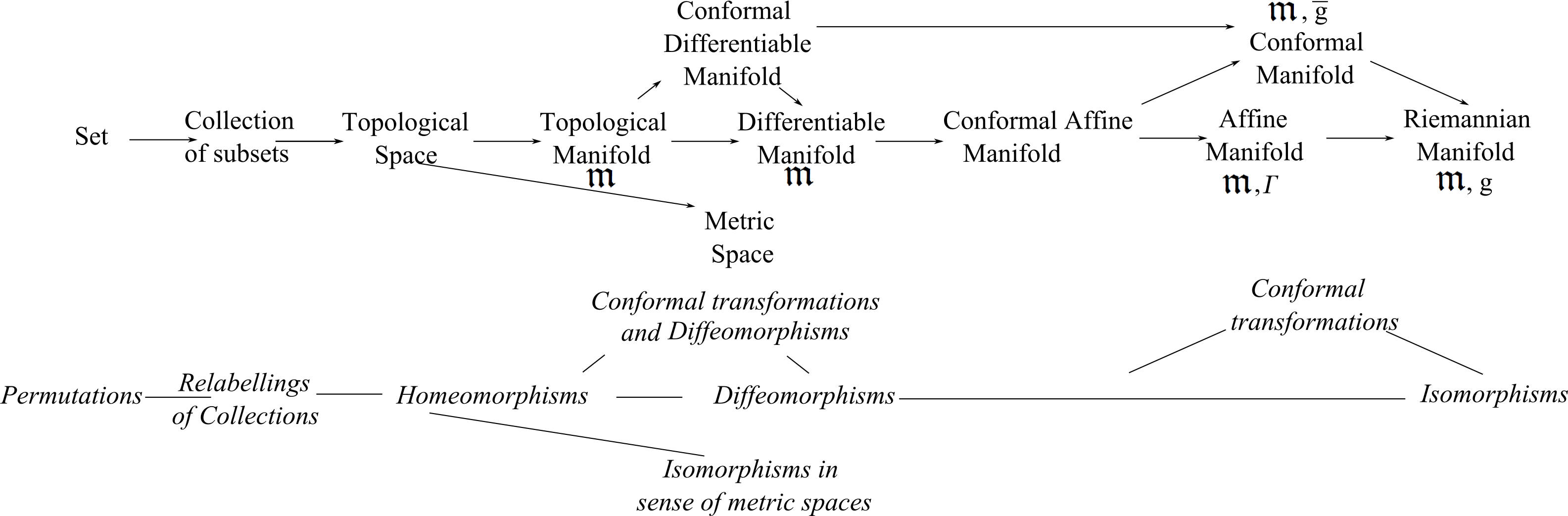}
\caption[Text der im Bilderverzeichnis auftaucht]{        \footnotesize{The levels of structure, now with corresponding morphisms indicated.} }
\label{Bigger-Set-2c} \end{figure}          }
%
\section{The corresponding spaces of spaces}

\subsection{Further classification}

Let the lower layers be dynamical, which subsequently translates to them then being quantized.
Two approaches to this are as follows. 

\mbox{ } 

\ni {\bf Program A)} Take each layer piecemeal.  
This might be interpreted as losing the upper layers to focus on the dynamical nature of the new topmost layer.

\mbox{ } 

\ni {\bf Program B)} Consider a range of layers.
E.g. retain the upper layers whilst letting the lower layers share in being dynamical (though on some occasions some upper layers are removed out of being in fact physically redundant).  

\mbox{ }  

\ni The idea then is to use the previous Sec and the distinction between Programs A) and B) to make a descent along Fig \ref{Bigger-Set-2}'s diagram of levels of structure based on sets.

\subsection{The usual upper levels' spaces of spaces}

As a starting point for classical configuration spaces, I already presented Riem($\bupSigma$) in Sec 1. 
This is well known in the Theoretical Physics literature, see e.g. \cite{DeWitt, Fischer70, Giu09} and \cite{ADM, I93, KieferBook} for the corresponding dynamics.  

\mbox{ } 

\ni Program A) As regards the space of differentiable structures on a given 3-space $\bupSigma$, one might well first wonder whether the amount of differentiability involved matters.
Differentiable structure for relativistic theories is usually taken to be ${\cal C}^{\infty}$ 
(in distinction from the analytic functions ${\cal C}^{\omega}$, since their analytic continuation precludes incorporation of causality).
Moreover, weakening ${\cal C}^{\infty}$ to ${\cal C}^k$ $k \geq 1$ makes little difference, 
by Whitney's demonstration \cite{Whitney36} that each such differentiable structure is uniquely smoothable. 
%
%
%
Also for dimension $< 4$ (including in particular the usual spatial dimension of 3), differentiable structure on a given topological manifold is unique.\footnote{This issue  
becomes nontrivial in higher dimensions, especially in dimension 4.}
%
Thus for the usual case of 3-$d$ space in GR, the space of differentiable structures per given topological manifold is trivial. 
Thus also this is not an additional layer of structure to build into the below approaches which transcend the differentiable structure level.  

\mbox{ } 

\ni Program B) What if all of the metric and differentiable structure (and affine structure, if independent) are dynamical?  
Then Superspace($\bupSigma$) applies again, and the counterpart with conformal differentiable structures are tied to CS($\bupSigma$).  
What Program B) here does nontrivially involve is `differentiable-level information on a given topological manifold $\bupSigma$ equipped with 3-metrics' 
alias `3-geometry' in Wheeler's parlance. 
The space of these spaces is indeed Superspace($\bupSigma$).

\mbox{ } 

\ni To discuss this further, it is important to note that quotienting (or reduction) kicks one out of the (infinite-$d$) manifolds into the (infinite-$d$) stratified manifolds.  
Thus we need to introduce these and it helps to start with the more fully understood finite mechanics case.
A simple example is the hemisphere with edge as the space of triangles in 3-$d$.  

\mbox{ } 

\ni {\bf Stratified manifolds} in general lose the three manifoldness axioms.
They are however still `nicely pieced together'.
They comprise a set of manifolds (not necessarily of the same dimension, so local Euclideanness is certainly lost), which 

\ni are none the less fitted together in a relatively benign manner by obeying Whitney's \cite{Whitney65} {\it frontier property}. 
Namely that, for any two of the manifolds, say $\FrMgen$, $\FrMgen^{\prime}$ with $\FrMgen \neq \FrMgen^{\prime}$
\beq
\mbox{ if } \FrMgen^{\prime} \bigcap \overline{\FrMgen} \neq \emptyset \mbox{ } , \mbox{ } 
\mbox{ then } \FrMgen^{\prime} \subset \overline{\FrMgen} \mbox{ and } \mbox{ } \mbox{dim}(\FrMgen^{\prime}) < \mbox{dim}(\FrMgen) \mbox{ } .
\eeq
[Here overline denotes closure.]
We then say a partition into manifolds has the {\it frontier property} if the set of manifolds has.
Finally, a {\it stratification} is a strict partition of one's topological space which has the frontier property.

\mbox{ }

\ni More generally, quotienting {\sl topological spaces} in general kicks one out of the restriction to Hausdorff and second countable topological spaces. 
However these properties are fortunately retained by many physically relevant examples of stratified configuration spaces; these are, in this sense, `2/3rds of a manifold'.
In these the stuctural differences are limited to different points being able to posses different kinds of neighbourhood, 
e.g. by variable dimension or the presence of boundaries, corners etc.

\mbox{ } 

\ni In particular, this case applies to Superspace($\bupSigma$) itself \cite{Battelle, Fischer70, Giu09}.
Here the strata correspond to different isotropy groups, i.e. what Killing vector structure each $\bupSigma, \bh$ possesses. 
This case also applies to the 3-body problem as an actual configuration space.  

\mbox{ } 

\ni Note 1) As regards the infinite-dimensionality complication, the manifold and stratified manifold concepts translate to \cite{Fischer70} cases in which $\mathbb{R}^n$ 
is replaced by suitable infinite-dimensional linear spaces.
This is well known for the case of Banach spaces \cite{AMP}, 
though the study of Superspace($\bupSigma$) itself makes use of a more general kind of Fr\'{e}chet spaces \cite{Fischer70, Giu09, Hamilton82}.  

\ni Note 2) Considering the conformal versions of the above turns out to be relatively straightforward, due to the contractibility of the conformal group \cite{FM96}.
By this, conformalized versions of GR's configuration spaces share topological properties with the un-conformalized versions.

\ni Note 3) GR connventionally retains a solitary global degree of freedom alongside the conformal 3-geometries \cite{York72}.
In the particle model counterpart (See Fig 5), adjoining a such corresponds to coning shape space back to relational space -- no global versus local scale variable distinction -- 
but for GR the two are clearly different.
The dynamical questions then are what is a pure conformogeometrodynamics, 
and what is a conformogeometrodynamics interacting with a solitary volume degree of freedom \cite{York72, York74, York7173ABFOABFKOShapeDyn}. 

\ni Note 4) Moreover, both Superspace($\bupSigma$) and CS($\bupSigma$) have sub-cases that are orbifolds and even manifolds \cite{FM96}.
This occurs for cases in which $\bupSigma$ does not admit any $\bh$ with symmetries.

\subsection{Affine, Connection and Loop variants}

{            \begin{figure}[ht]
\centering
\includegraphics[width=0.7\textwidth]{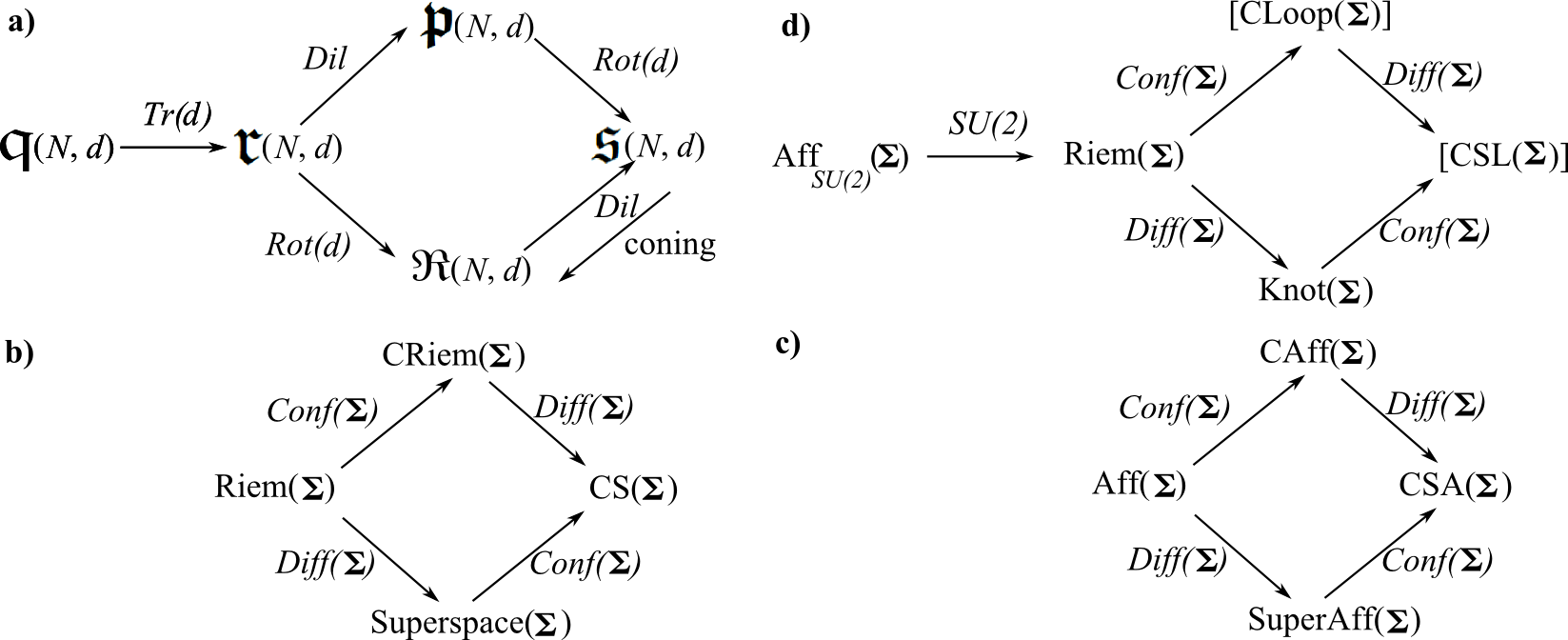}
\caption[Text der im Bilderverzeichnis auftaucht]{        \footnotesize{a) RPM configuration spaces, now including the coning condition.
b) GR-as-geometrodynamics's configuration spaces.
c) Affine theory and d) loop formulation of GR's configuration spaces. 
The last of these introduces an extra local $SU(2)$ in defining is variables.   
Taking out these degrees of freedom, one passes to the space of loops and  subsequently quotienting out Diff'$\bupSigma$) as well, to the knots. 
Conformal versions are not usually studied in this `Loop Quantum Gravity' case; for further study of spaces of spaces here, see e.g. \cite{GPBookAL93Gielen}.  } }
\label{RPM-Riem-Aff-Loop} \end{figure}          }

\ni Program A) If only affine connections contain physically meaningful information, so $\FrQ$ = Aff($\bupSigma$), what dynamics do these support?
If only `Weyl connections' contain physically meaningful information, so $\FrQ$ = CAff($\bupSigma$), what dynamics do these support?


\ni Program B) What if the spatial affine connection is dynamical independently from the metric? 
I.e. $\FrQ$ = Riem($\bupSigma$) $\times$ Aff($\bupSigma$). 
[Vary these separately, rather than presupposing that the affine connection present is the metric connection.] 
Here I also note that Schreiber \cite{Schreiber} has recently reviewed the modelling of observables by sheaves.

\subsection{The Topological Manifold level}\label{TopMan}

\ni Note that conformal mathematics -- conformal field theory (CFT), conformal geometry -- 
and supersymmetric mathematics are the first and most straightforward variants of the most standard fundamental theories of Physics.
Topological variants are the next most straightforward: topological field theory (TFT) \cite{TFT}, topology change in GR, one level of structure further down. 
Considering levels yet further down than that is rare so far in the Theoretical Physics literature, Isham's own work being a notable exception. 

\mbox{ }  
 
\ni In standard geometrodynamics, one firstly assumes a fixed spacetime topological manifold; this is a major restriction on possible spacetime-level solutions. 
Secondly, one assumes a fixed $\bupSigma \times \mathbb{R}$ is a subset of this restriction that accommodates the standard geometrodynamical and canonical approaches.  
Some further restrictions to certain spatial topologies are sometimes motivated or used are as follows.

\mbox{ } 

\ni i)   The orientable $\bupSigma$.

\ni ii)  The connected $\bupSigma$.

\ni iii) The compact without boundary $\bupSigma$.     

\ni iv)  Whether to append the singular spaces.  
These do not have the chart property, even if the topological manifold change is elsewise `tame'  (e.g. dimension preserving and/or closedness preserving). 

\mbox{ } 

{            \begin{figure}[ht]
\centering
\includegraphics[width=1.0\textwidth]{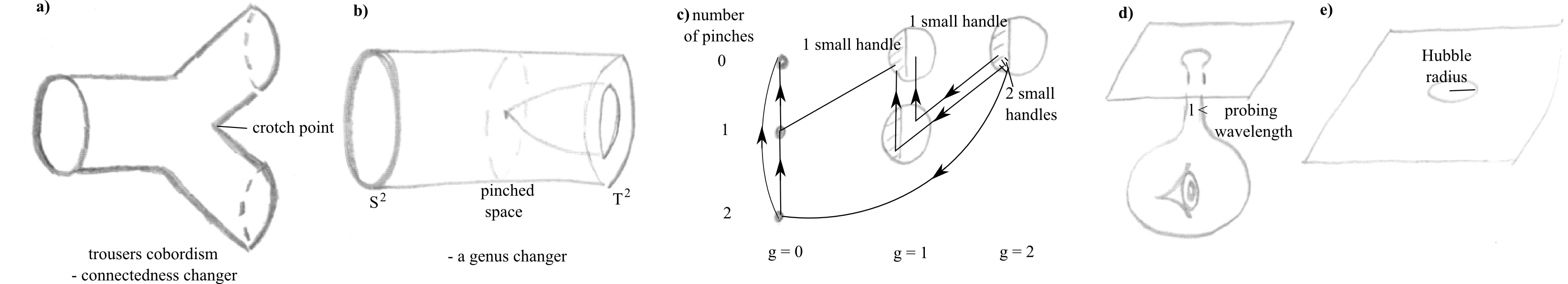}
\caption[Text der im Bilderverzeichnis auftaucht]{        \footnotesize{a) and b) Examples of cobordisms. 
c) Depiction of what is near what in BigRiem in the simple case of 2-$d$ orientable connected compact without boundary $\bupSigma$.  
d) Open universe that looks closed, by use of one of Hawking's tubes of negligible action. 
e) A closed universe that looks open.
d) begs the further question of whether such tubes are dynamically stable.} }
\label{Big-Superspace}\end{figure}            }

\ni What happens if each of these assumptions is lifted? 
The idea of working at this level was first put forward in Wheeler's ideas of `spacetime foam' \cite{Wheeler57, WheelerGRT, Battelle}.  
This was motivated from the form of the Feynman path integral formulation: generalization from Feynman diagrams to topology change diagrams... 

\mbox{ }

\ni As a first part of answering this, Geroch's Theorem \cite{Geroch66} requires one to choose between Scylla and Charybdis: singularities on the one hand, or a loss of causality on the other.  
This result is based on a combination of mathematics pioneered by Milner and Thom: singular metric geometries, cobordisms and Morse Theory.
At the classical level, one often chooses the former, and thus allows for the inclusion of singular metrics.   
%

\mbox{ } 

\ni {\bf Cobordism}. $\Sigma_1$ and $\Sigma_2$ are cobordant if there exists an M of dimension one greater which interpolates between them (Fig 6.a-b).  
One can view the piece in between as $\Sigma \times [0, 1]$ with some parameter $\lambda$ ranging over [0, 1].  
To be clear, the $\FrT$ group is {\sl not} a group of the conceptually simpler kinds that occur in topology -- homotopy, homology or cohomology: {\sl characterizers of} topology.  
It is, rather, a `ripping group' of topology-{\sl altering} operations. 
E.g. for 2-$d$ orientable closed without boundary manifolds, cobordisms are generated by the genus-changing operation of `adding handles'.  
For instance, the operations of {\it Surgery Theory} are rippings.  

\mbox{ } 

\ni {\bf Morse Theory} has ties to critical points and to cohomology.
The part of this that we require for this article are {\it Morse functions} $f$. 
These are characterized by their Hessians at each critical point; these critical points are isolated and nondegenerate for $f$ Morse.  

\mbox{ } 

\ni Then addressing this level's Program A), TopMan is the space of $\bupSigma$'s themselves.
This is possibly subject to some restrictions from a menu such as a particular dimension, orientable, compact, connected.
What dynamics do these support? 
E.g. a field theory with only TFT degrees of freedom.

\mbox{ } 

\ni To address Program B), indeed the most commonly encountered Program B) involves the metric and toplogical manifold levels.
This is `geometrodynamics with topology change', which usually involves a collection of classical impasses motivated by the corresponding quantum path integral problem involving 
`transition amplitudes for topology change'.

\mbox{ } 

\ni Firstly arm each with whatever differentiable structures and metrics that these can take.

\mbox{ } 

\ni However, this is still dynamically sterile through not providing suitable intermediates: Riem($\bupSigma$) for $\bupSigma$ now singular.  
One gets around this by adjoining pinched manifolds also.
Then we have BigRiem (Fig 6.c), BigSuperspace \cite{Fischer70} etc, on the set of ordinary and pinched $\bupSigma$.  
An example of this is a geometrodynamics that is also a topolodynamics in the sense of admitting changes in topological manifold.
In this regard, Fischer \cite{Fischer70} comments that: {\it ``nevertheless, it is hoped that the complete Superspace of all possible topologies can be pieced together 
from the individual superspaces, so that quantum fluctuations in the topology can be described.  
This I believe will be possible only after a great deal is known about the individual superspaces."}
He added that {\it ``In such a program, pinched manifolds will play the crucial intermediary role in the topology change."} 

\ni As a simple case, for compact 2-$d$ oriented manifolds, genus is the sole characterizer (Fig 6.c).
Consider furthermore just the spheres, tori and the pinched spheres that lie between these.
This requires studying the counterparts of Riem($\bupSigma$) and Superspace($\bupSigma$) for the pinched spheres, 
and then the BigRiem and BigSuperspace on the three topologies in question.

\mbox{ } 

\ni As regards the issue of {\sl how} singular these pinched spacetimes are to be, {\bf Morse spacetimes} \cite{Top-Change} are a nicely tame example.  
Morse Theory encodes these metrics singularness via a Morse function: $f:\FrM \rightarrow [0, 1]$ with $f^{-1}(0) = \bupSigma_1$ and $f^{-1}(1) = \bupSigma_2$ 
and $r$ critical points in the interior of $\FrM$.  
These are isolated and nondegenerate for $f$ a Morse function.  
So then one is more specifically interested in e.g. BigRiem(Morse), BigSuperspace(Morse) and so on.

\subsection{Metric Space level} 

Program A) For $\mX_1$ and $\mX_2$ subsets of a metric space ($\mM$, Dist), the {\it Hausdorff distance} is
\beq
\mbox{Dist}_{\sH}(\mX_1, \mX_2) = 
\mbox{max} \left(\sup_{x_1 \in \sX_1}  \inf_{x_2 \in \sX_2} \mbox{Dist}(x_1, x_2), \, \sup_{x_2 \in \sX_2} \inf_{x_1 \in \sX_1} \mbox{Dist}(x_1, x_2) \, \right) 
\eeq
Then the {\it Gromov--Hausdorff} distance between any two compact metric spaces $\mM_1$ and $\mM_2$ is is 
\beq
\mbox{Dist}_{\sG\sH}(\mM_1, \mM_2) = \mbox{inf}\big(\mbox{Dist}_{\sH}(f_1(\mM_1 ), f_2(\mM_2 )\big)
\eeq
over all isometric embeddings $f_1: \mM_1 \rightarrow \mM$ into all metric spaces $\mM$.
[The embeddings render $\mM_1$ and $\mM_2$ into the form of subsets of larger metric spaces within which the Hausdorff distance notion applies.]
Thus we have an example of a space of metric spaces which is itself a metric space.  

\mbox{ }

\ni Program B) As regards how the (positive-definite) Riemannian manifolds sit within the metric spaces, each Riemannian metric induces a metric in the metric space sense: 
the {\it path metric} 
\beq
\mbox{Dist}(x, y) = \inf_{\mbox{\scriptsize $\gamma \mbox{ from } x \mbox{ to } y$}}\int_{\gamma} \sqrt{\mg_{\mu\nu}(x)\d x^{\mu}\d x^{\nu}} \mbox{ } .
\eeq
Then supposing only metric spaces contain physically meaningful information, what dynamics do these support?

\subsection{Topological Space level} 

\ni Program A) As regards space of topological spaces, Top($\mX$) on a fixed set $\mX$ form a lattice \cite{Lattice}.  

\mbox{ }

\ni To understand lattices in general, introduce first a {\it poset} (partially ordered set). 
I.e. a set $\mX$ equipped with a {\it partial order}: a reflexive, antisymmetric and transitive binary relation (e.g. $\subset$, or $\geq$). 
Then a {\bf lattice} is then a poset within which each pair of elements has a least upper bound and a greatest lower bound.
In the context of a lattice, these are called {\it join} $\lor$ and {\it meet} $\land$.  
$\lor$ and $\land$ form an algebra. 
Each of these operations is idempotent, commutative and associative, and the pair of them obey the {\it absorption conditions} $a \lor \{ a \land b\} = a$  and $a \land \{a \lor b\} = a$.

\mbox{ }

\ni For the lattice of topologies on a fixed set, the partial order in question is the relative coarseness of the topologies in question. 
This particular lattice is complete and complemented \cite{Lattice}; \cite{I89-Latt} lists further properties.
%
%

\mbox{ } 

\ni Then supposing only topological spaces contain physically meaningful information, what dynamics do these support?

\mbox{ } 

\ni Program B) What if all of the topological space and any subsequent emergent differentiable, affine, conformal, and metric structure are dynamical?
Now, far from all topological spaces support topological manifolds, so one encounters a problem unless one can regard the structures from there upward as emergent for certain $\tau(\mX)$. 
This does however beg the question of what {\sl other} structures some topological spaces are capable of supporting.
Thus this is a major breakdown point: the tower versus the far wider ranges of topological spaces and of collections more generally (Fig \ref{Collections-and-Tower}).  
A dynamical theory of topological spaces might explain whether and how topological manifolds are prevalent in the set of possible universes as a zeroth principles theory. 
After that, one would restrict attention to a TopMan($\mX$) first-principles theory.


\ni Further discussion in this article involves a number of particular useful classes.

\mbox{ } 

\ni 1) The Hausdorff second-countable locally Euclidean case of this are of course just the {\sl manifolds} again.

\ni 2) Their generalization to Hausdorff second-countable topological spaces. 

\ni 3) Aside from the manifolds, another subcase of these significance below are the Hausdorff second-countable locally compact topological manifolds (H2LC).  

\ni 4) The further generalization to the {\sl stratified manifolds}.
 
\ni 5) The {\sl Hausdorff second-countable stratified manifolds} subcase of 2) have already been mentioned as a useful class with `2/3 of the manifold properties'.

\ni 6) The combination -- {\sl H2LC stratified manifolds} is the one that enters the {\it stratifold} construct \cite{Kreck};
these are a particularly well-behaved class as far as stratified manifolds go.    
Since locally compact Hausdorff implies compact, this includes a number of physically relevant stratified manifolds.

{            \begin{figure}[ht]
\centering
\includegraphics[width=0.6\textwidth]{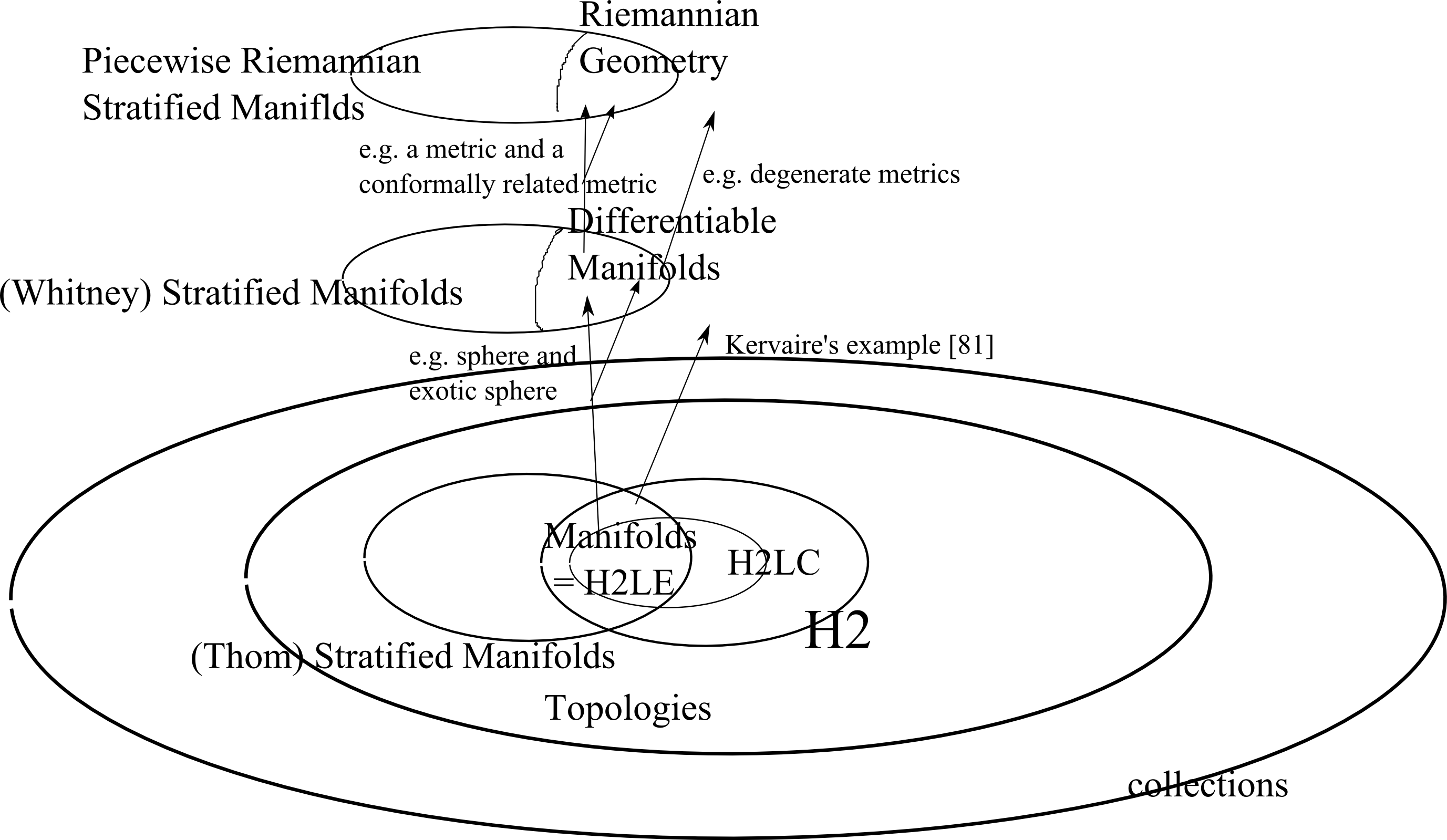}
\caption[Text der im Bilderverzeichnis auftaucht]{        \footnotesize{The collections and the tower. 
It illustrates by examples that not all cases within a given floor of the tower can be extended up the tower.} }
\label{Collections-and-Tower}\end{figure}    }

\subsection{Collection of Subsets Level}

\ni Program A) If only collections of subsets of sets contain physically meaningful information, so $\FrQ$ = Collect($\mX$), what dynamics do these support?
The issue here would be minus oneth principles: 
can it be shown that consideration of dynamical collections of subsets has a propensity to lead to collections that are specifically topologies?

\mbox{ } 

\ni Program B) What is effect on the upper layers of structure if `the underlying set' is allowed to quantum-mechanically fluctuate? 
%

\subsection{Sets Level}

Program A) If only sets contain physically meaningful information, so if $\FrQ$ is the set of sets, what dynamics do these support?

\mbox{ } 

\ni Program B) leads to problems since the space of topological spaces becomes unruly if one tries to define it on the set of sets rather than on a fixed set $\mX$.  
This is rooted in the set of sets suffers from Russell's paradox, limiting the study that far of spaces of space \cite{I89-Latt}.  
Yet it is conceivable -- if the universe is built out of sets -- that all of the metric, connection, differentiable structure, topological manifold, topological space and set 
are dynamical (or, a fortiori, undergo something like quantum fluctuations).

\subsection{A first commentary on other paradigms of mathematics}\label{Commentary}

In the equipped sets paradigm, sets are viewed as the primary entities to equip.
One alternative to this is to replace what is regarded as primary entities by such as categories or topoi.
See Sec \ref{Last} for another type of alternative.

\mbox{ }

\ni{\bf Categories} \textgoth{C} = (\textgoth{O}, \textgoth{M}) consist of objects \textgoth{O} and {\it morphisms} \textgoth{M} (the maps between the objects, 
\textgoth{M}: \textgoth{O} $\longrightarrow$ \textgoth{O}, obeying the axioms of domain and codomain assignment, identity relations, associativity 
relations and book-keeping relations. 

\ni {\bf Functors} are then maps \textgoth{\Large F}: \textgoth{C}$_1$ $\longrightarrow$ \textgoth{C}$_2$ that obey various further 
axioms concerning domain, codomain, identity and action on composite morphisms.  

\ni A {\it small category} is one for which both \textgoth{O} and \textgoth{M} are sets. 

\ni A {\it locally small category} is one for which for each \textgoth{O}$_1$ and \textgoth{O}$_2$, \textgoth{M}(\textgoth{O}$_1$ and \textgoth{O}$_2$) are a set. 

\mbox{ } 

\ni See e.g. \cite{MacLane} for more details.
One can then reframe the question of spaces of spaces for Theoretical Physics in terms of spaces of categories or indeed categories of categories.  
Here I note that while the set of sets impasse carries over to the category of categories, it does not carry over to the small categories, 
since the category of small categories is not itself small. 
Finally, one consider all of dynamical evolution, probability theory and quantization on categories more generally (Isham pioneered the last of these \cite{I03}).   

\mbox{ } 

\ni For a brief outline of topoi, see Sec \ref{Last} since to date these have mostly only been used at the quantum level.

\section{Background Independence and the PoT for the deeper levels}\label{Deeper}

\subsection{Temporal and Configurational Relationalism apply to all levels}\label{Rel-All}

\ni 1) {\bf Temporal Relationalism}.  
One can consider an absense of extraneous times or time-like variables at all levels, and likewise as regards label times being meaningless.
If a theoretical framework has notions of generalized configuration $Q^{\sfA}$ and change of generalized configuration $\d Q^{\sfA}$, 
it also has a notion of action, $S[Q^{\sfA}, \d Q^{\sfA}]$  
It consequently has a notion of generalized momentum (momentum can be defined in terms of change rather than velocity \cite{FileR, TRiPoD}).
In Isham's categorical version \cite{I03}, generalized configurations are objects and generalized momenta the corresponding arrows alias morphisms.
Then in such settings, a notion of generalized constraint continues to make sense,
and Dirac's argument that reparametrization invariance enforces at least one primary constraint continues to apply also.
Thus reparametrization invariance or some modified successor notion that makes no reference to the physically meaningless parameter gives rise to a generalized 
$\scC\scH\scR\scO\scN\scO\scS$.

\mbox{ } 

\ni Moreover, one can consider changes of structure at whichever level, 
upon which one can base a Mach's time principle resolution and its more specific STLRC implementation.
Rearrangement of the generalized $\scC\scH\scR\scO\scN\scO\scS$ gives an emergent time expression along the lines of (\ref{t-em-Mach}).

\mbox{ } 
 
\ni 2) {\bf Configurational Relationalism}.
The notion of a group of physically irrelevant transformations $\FrG$ also makes sense at all levels of mathematical structure.
In particular this could be (some subgroup of) the automorphism group of the generalized $\FrQ$, $Aut(\FrQ)$.  
Then one is to pass to the corresponding quotient, e.g. [compare (1)]
\beq
\tilde{\FrQ} = \FrQ/Aut(\FrQ)
\label{Aut-Rel}
\eeq 
in the case of the full $Aut(\FrQ)$.  
One can read these quotients off Fig 4.  
Note that whilst Isham said the PoT mostly concerns diffeomorphisms \cite{I93}, this is implicitly within the context of metric to differentiable structure level.
This is clear since he evokes the $\mX$/$Perm$($\mX$) quotient in \cite{I89-Latt, I89-Rev} as a direct analogue of 
                                       Wheeler's quotient Superspace($\bupSigma$) = Riem($\bupSigma$)/$Diff$($\bupSigma$).
One can take Isham's (\ref{Aut-Rel}) as a generalization of Kendall--Barbour shape theory and its scaled counterpart, and of the the GR analogues of both.

\mbox{ } 

\ni The indirect $\FrG$--act $\FrG$--all implementation also extends to all levels.
This applies firstly in the single-layer context of Program A), in the form (\ref{G-Act-G-All}) with $\FrG$ generalized to apply at that particular level.
Secondly in the multi-layer context of Program B)
by use of 
\be
\mbox{\Large S}_{g_1 \in \sFrG_1} 
\mbox{\Large S}_{g_2 \,\, \si\sn \,\, \se\sa\scc\sh \sFrG_2 \,\, \scc\so\sr\sr\se\sss\sp\so\sn\sd\si\sn\sg \, \, \st\so \, \,  \st\sh\sa\st g_1}
\circ \mbox{ Maps } \circ \stackrel{\rightarrow}{\FrG}_{g_2} \circ \mbox{ Maps } \stackrel{\rightarrow}{\FrG}_{g_1} O \mbox{ } .
\ee
See e.g. Sec \ref{BI-Top-Man} for a commonly encountered example of this.  

\mbox{ } 

\ni As regards types of $\mbox{\Large S}_{g \in \sFrG} $, as one descends the levels one needs to restrict to discrete versions: sum, discrete average, inf and sup.
(Discrete) action constructs give rise to $\scS\scH\scU\scF\scF\scL\scE_{\sg} = 0$ constraints in association with the procedure in Fig 9.a).

\subsection{Canonical and brackets structure level by level}\label{Can-All}

Allot now also corresponding generalized classical brackets.  
Then the notions of brackets algebraic structure formed by constraints (\ref{CC}), 
Dirac procedure and of quantities forming some kind of zero brackets with constraints continue to make sense.
It is more straightforward to envisage that, once at the quantum level, commutator brackets persevere.
So ultimately, at least if something like quantum theory remains pertinent at these levels, 
3) {\bf Constraint Closure} and 4) {\bf Expression in terms of Beables} transcend all the way down the levels too.
However, the linear--quadratic (or more generally linear--nonlinear) distinction may no longer apply, 
without which the partition of the Dirac beables concept (\ref{D-B}) by the \K beables concept (\ref{K-B}) would not carry through.

\subsection{Spacetime--space distinctions level by level}\label{Spacetime-All}

There are distinct spacetime and space floors for each of the levels of structure (Fig 8). 
Some (but not all) of the features which distinguish spacetime from space are progressively lost at the deeper levels of mathematical structure.
Signature is lost, then codimension 1 becomes meaningless with loss of dimension, leaving space being a strict subset of spacetime.

{            \begin{figure}[ht]
\centering
\includegraphics[width=0.75\textwidth]{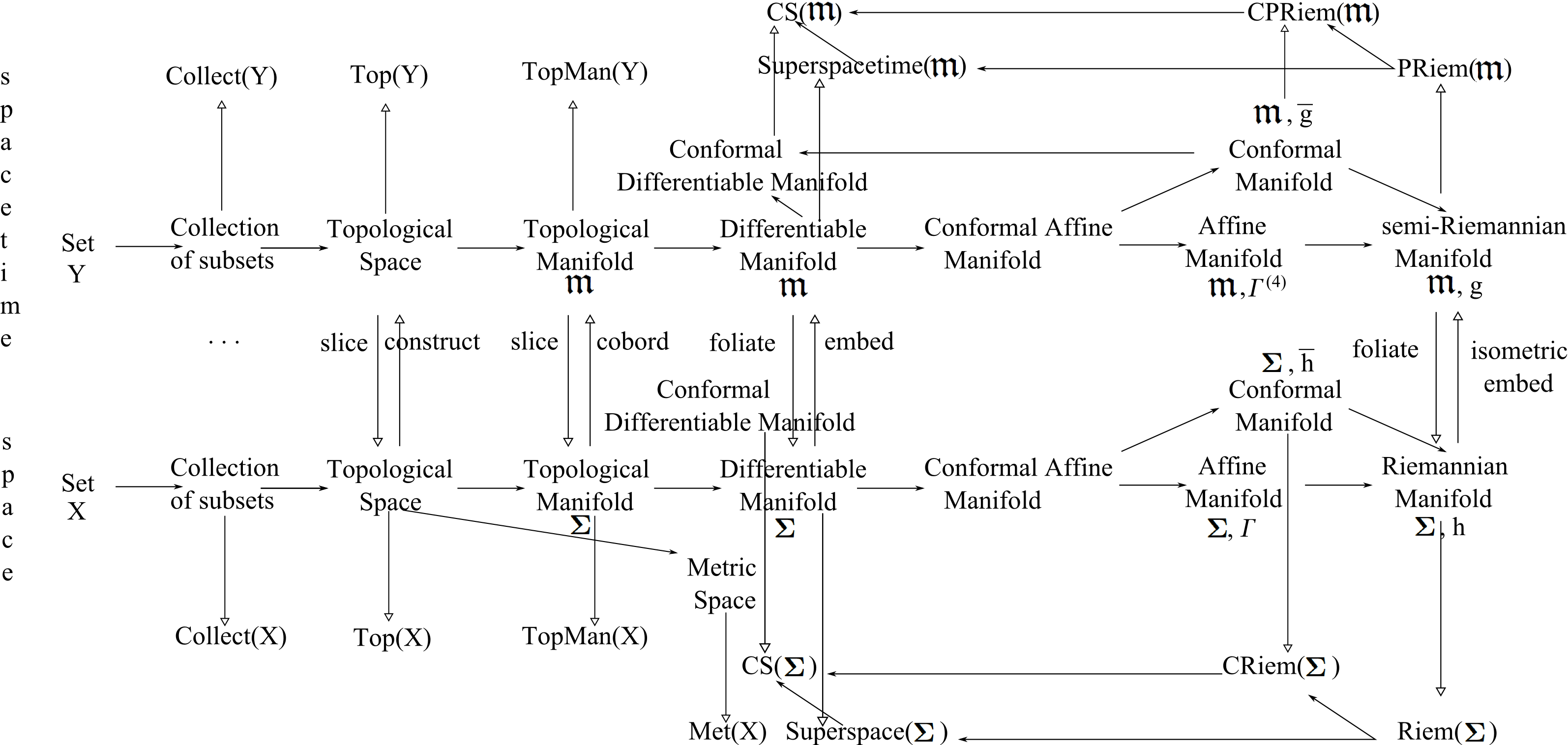}
\caption[Text der im Bilderverzeichnis auftaucht]{        \footnotesize{Specific spacetime and space versions, interlinked with maps and including corresponding spaces of spaces.
Note in particular the advent of stratified manifolds.} }
\label{Bigger-Set-7}\end{figure}            }

\ni For instance, signature-dependent differences are lost beneath the level of conformal metric structure.
However, some distinction remains. 
This is due firstly to overall dimension remaining a meaningful concept down to the level of topological manifolds (by to their local Euclideanness).
Thus one has e.g. a 3-$d$ entity and a 4-$d$ entity of the same type in the role of spacetime.
Secondly, even beneath that level, there are still distinct larger and smaller entities, e.g. a `space slice' subset of a `spacetime' set. 
This still retains some meaningful identity even in structurally sparse conditions, such as being an antichain within a poset in the Causal Sets Approach \cite{Sor91, Sor97}.  
%

\mbox{ }

\ni Thus the idea of spacetime (plus matter fields defined upon it) versus spatial configuration (plus field configurations) as distinct starting points prevails to all levels.
Consequently so does the consideration of a group of physically irrelevant transformations $\FrG_{\sS}$ specifically for the former: 5) {\bf Spacetime Relationalism}.  

\ni \beq
\mbox{Superspacetime}(\FrM) := \mbox{PRiem}(\FrM)/Diff(\FrM) 
\eeq
-- the space of spacetimes modulo spacetime diffeomorphisms -- is not Hausdorff (\cite{Fischer70} reported that Stern also proved this result).  
$\FrG_{\sS}$'s generators continue to require closure and to have an associated notion of observable.

\mbox{ }

\ni Finally, can also imagine histories and timeless records for whatever level of mathematical structure assumed, including subjected to groups of physically irrelevant transformations.

\subsection{Slicing and constructing to all levels}

The `two-way passage between' spacetime and space issue carries through, as follows.  

\mbox{ }

\ni 6) {\bf Slice Independence} is the level-independent extension of Foliation Independence. 
If this does not occur, one has a {\it Slice Dependence Problem}

\ni 7) {\it Reslicing Invariance} would be the level-independent generalization of Refoliation Invariance, which itself is only meaningful as far as the Topological Manifold level.   
We do not however know if this holds for the general level.  
Thus I pose the question 
-- which Teitelboim affirmed \cite{T73} using the metric and differentiable manifold levels' Dirac algebroid -- {\sl at all subsequent levels of mathematical structure}. 

\mbox{ }

\ni {\bf Spacetime Construction} is a valid issue to investigate at each level to the extent that each level has spacetime to reconstruct. 
That applies both to a) reconstruction by going up levels and b) to reconstruction within each level from its space notion to its spacetime notion.
b) is in a sense an inverse procedure to slicing up the spacetime notion into a sequence of space notions.  
However, it is a harder inverse since it assumes only the `spatial' structure whereas the slicing move can assume the entirety of the level in question's `spacetime' structure.  

\mbox{ } 

\ni All in all, I use `space', `time', `spacetime', `slice', `foliate', `surround' and `construct' as level-independent concepts (as per Fig 9.b).  
{\sl This conceptualization indeed points to many further versions of the PoT facets at each of these levels of structure.} 

\mbox{ } 

\ni Finally, at any level of structure, it is desirable for timefunctions and all other steps involved in the above exposition to be 8) {\bf Globally Valid}.

{            \begin{figure}[ht]
\centering
\includegraphics[width=0.5\textwidth]{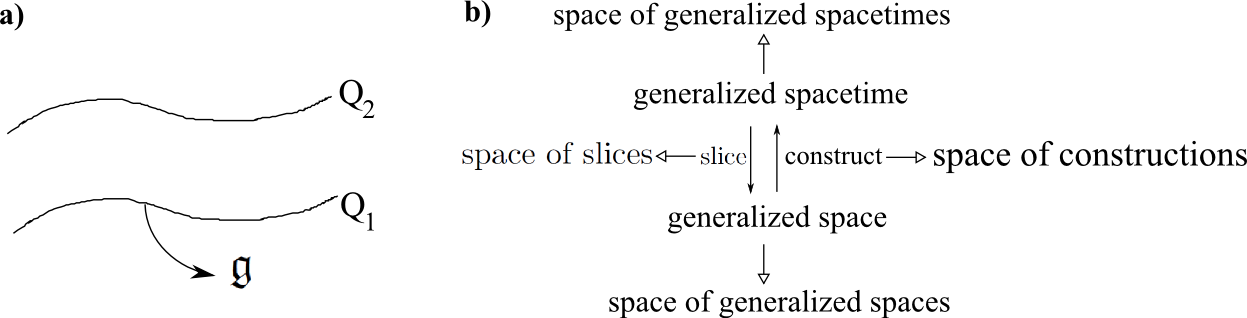}
\caption[Text der im Bilderverzeichnis auftaucht]{        \footnotesize{a) Generalized Configurational Relationlism.
b) The generalized eightfold cross.} }
\label{Fig-9}\end{figure}    }

\section{Level-dependent classical issues}

\subsection{Some Background Independence and time aspects lost in earlier stages of the descent}

Firstly, Newtonian theory and SR involve types of metric background dependence.  
Secondly, Rot($d$), the Euclidean group Eucl($d$), the Lorentz group SO($d$, 1) and the Poincar\'{e} group Poin($d$) are metric-level structures because they are {\sl isometry} morphisms.
Thirdly, spacetime signature enters a number of aspects of SR and GR (including allotting a GR coordinate time).  
This is a bit more than just a metric-level notion since there is still distinction between spacelike and indefinite geometries (at the differentiable manifold level), 
but it does not further transcend to the topological level.
%
%
%
One needs relativistic causality in order to consider a) time-orientability.
b) Closed timelike curves (including the discarding of spacetimes containing these).
c) Foliations with respect to spacelike surfaces that represent sequences of instants of time as experienced by arbitrarily moving families of observers.  

\mbox{ } 

\ni On the other hand, differentiable and affine manifolds upon which a metric is not defined do not carry space--time distinction or relativistic notions of causality. 
%
%
In a wider range of contexts in which time is regarded as a coordinate or parameter, 
it is to additionally be reparametrizable not only by $A + Bt$ but also by any transformation that respects monotonicity.  
Taking time to be a {\it monotonic} (rather than direction-reversing) function itself is meaningful down to topological spaces using continuous functions.
Below that it can still be monotonic as regards the surviving ordering property of time notions.
Diffeomorphisms are the differentiable manifold level morphisms, and no more.
Finally note that clocks are {\it material} entities, though some of their general features might be expected to still exist in extreme de-layered regimes of the universe.

\subsection{More detailed limitations of relevance to the first four facets}

We have argued that if a theoretical framework has notions of configuration $Q^{\sfA}$ and change of configuration $\d Q^{\sfA}$, it has a notion of action.
Note that sufficient continuity and differentiability is required to define the usual kind of action's integral and for the standard Calculus of Variations to subsequently apply. 
These cease respectively at the topological space level and somewhere around the differentiable manifolds level 
(stratified differentiable manifolds retain differentiability in some neighbourhoods).
For instance, defining an action sufficiently far down the levels requires use of a discrete sum rather than a continuous integral (e.g. not all topological spaces are measurable).
One also requires a discrete counterpart of the definition of momentum and of the Calculus of Variations in order to operate beyond the level at which differentiability ceases. 
This leads to discrete analogues of the Euler--Lagrange equations and so on.  
(See e.g. \cite{DH13} for an example of a worked out treatment of discrete Calculus of Variations including constraints.)  

\mbox{ } 

\ni Also, the metric level's motivation to keep such actions at most quadratic may cease to apply at the lower levels 
(though the notion of decoupling variables and Ostrogradsky type reduction continue to exist withing discrete counterparts).
This has the knock-on effect of $\scC\scH\scR\scO\scN\scO\scS$ not necessarily being quadratic like GR's $\scH$ was.

\subsection{Relationalism and Background Independence at the level of topological manifolds}\label{BI-Top-Man}

\ni 1) {\it Temporal Relationalism}.
One can pass from cobordism in terms of a parameter to a reparametrization invariant formulation to a formulation without even reference to any inconsequential parametrization.
\ni $\int_{\tau} \d S(\tau, \d \tau)$ for the action of Program A) and $\int_{\tau}\int_{\sm \mbox{ for each } \tau} \d S(\tau, \mm, \d \tau, \d \mm)$ in Program B).  
Then $\tau$ has opportunity to enter the relevant changes.
[Here $\tau$ are topological manifold level variables and $\mm$ are metric geometry level ones.]

\mbox{ }

\ni 2) {\it Configurational Relationalism}.  
The Thin Sandwich Problem itself involves metrics and spatial hypersurfaces. 
Its generalization to Best Matching is a differentiable manifolds level construct via its use of the Lie derivative. 
Moreover, some of the concepts involved here, such as the Thin Sandwich's path integral analogy and the idea of matching pairs of configurations, 
continue to apply at the topological level and below.

\mbox{ } 

\ni The $\FrG$-act, $\FrG$-all method is fully general from the perspective of levels of mathematical structure.  
For Program A), this involves using 
\be
\mbox{\Large S}_{\tau \in \tFrT} \circ \mbox{Maps } \circ \stackrel{\rightarrow}{\FrT}O \mbox{ } .
\ee
On the other hand, the indirect approach to Program B) involves using
\be
\mbox{\Large S}_{\tau \in \tFrT} \mbox{\Large S}_{g \in \sFrG} \circ \mbox{Maps } \circ \stackrel{\rightarrow}{\FrG} \circ \mbox{ Maps } \stackrel{\rightarrow}{\FrT}O \mbox{ } .
\ee
such as a sum over all metrics over all topologies, meaning topological manifolds and often restricted to e.g. the compact connected ones.  
This is in the sense that given a particular topology, one can establish a $\FrG$-corrected version of $\FrQ$ on it. 
Here one often restricts to fixed dimension.
This corresponds to BigRiem =$ \bigcup_{\sbSigma}$Riem($\bupSigma$). 
This is often still further restricted to some combination of Sec \ref{TopMan}'s assumptions i) to iv) as regards $\bupSigma$. 


\ni Note 1) In the simple example of connected, compact without boundary oriented 2-$d$ manifolds, 
one is reduced to $\mbox{\Large S}$'s indexing set being just over the discrete genus parameter.

\ni Note 2) Though extremization or taking the inf or sup looks strange in that it could well pick but one contributor.  
It is far from clear whether it would make any sense to assume the universe has three holes just because the action for a three-holed 
universe is {\sl somewhat} larger than that for any other number of holes! 
See however Sec \ref{QBI} for a standard QM amelioration of this issue.  

\ni Note 3) I also point out an intermediate case: consider Program A) but with its own nontrivial $\FrG$ = Homeo($\bupSigma$): the topological manifolds up to homeomorphism.
This also has a 2-level formula in the sense of rippings of topological manifolds up to homeomorphism classes.  

\ni Note 4) As regards assumption iii) this is e.g. on Machian grounds or on grounds of simplicity (it certainly remains plausible that our universe is closed). 

\ni Note 5) Moreover that that in fact there is no observational basis for assuming that the universe's spatial topology is open or closed (Fig \ref{Big-Superspace}.d-e). 

\ni Note 6) Background Independence is more widely suggestive of seeing what happens if one lifts restictions i) to iv).

\mbox{ }

\ni Cobordism is a means of matching or comparing distinct topological manifolds.   
As regards the relation between spatial 3-topology and spacetime 4-topology, the topological analogue of 3-metric manifolds {\sl embed} into 4-metric manifolds 
involves 3-topologies that are cobordant to 4-topologies.
In the specific case of Program B) there is a further Euclidean versus Lorentzian cobordism distinction at this stage.  
The parameter involved in a Lorentzian cobordism might furthermore be thought of along the lines of a time coordinate.

\mbox{ }

\ni Finally, for Morse spacetimes, the Morse function $f$ also serves as a global timefunction for these spacetimes.
By possessing this, these spacetimes preclude time-nonorientability and the presence of closed timelike curves \cite{Top-Change}.

\subsection{Relationalism and Background Independence for metric and topological spaces}\label{Cl-Top-2}

The position-dependence of most notions of time in field theories carries over to such as spacetime lattices or space-continuum time-discrete versions.   
This also applies to evolution laws: such as difference-equation and stochastic versions of these exist in the absense of enough structure to do calculus.  
Also note that duration is a metric space but not topological space property.

\mbox{ } 

\ni Configuration spaces are generically stratified manifolds in reduced approaches, and these live `further down' the levels of mathematical structure than topological manifolds do. 
Thus Configurational Relationalism provides a distinct second reason to Isham's `quantize all the way down' as regards why to consider the layers of structure assumed. 

\mbox{ }

\ni At the level of metric spaces, some choices of $\FrG$ are id and the metric space isometries $Isom(\mX)$.  

\mbox{ } 

\ni The standard coordinate notion goes no further down than topological manifolds. 
Stratified manifolds still possess non-global notions of coordinates.
It is just that now charts differ in dimension and can also exhibit a generalization of the chart concept for a space with boundary.   

\mbox{ } 

\ni Each of the notions of orientability, curves and foliations do not go past the level of topological manifolds either 
(though to some extent they continue to be meaningful e.g. for stratified manifolds).  

\mbox{ }

\ni A further idea is {\it topologenesis} in the sense of topological manifolds arising from more general topological spaces.

\mbox{ }

\ni Program A) $\FrG$ application is a quotienting, which does not preserve a number of topological properties including the three that constitute manifoldness. 
Consider then what dynamics a topological-level stratified manifold can possess.

\mbox{ }

\ni Program B) Consider also what dynamics a topological through to metric level stratified manifold can possess.  

\mbox{ } 

\ni Finally consider whether one has a zeroth principle explaining stratified manifolds arising within a theory of dynamical

\ni topological spaces.
Some choices of $\FrG$ here are id and Homeo($\FrMgen$).  

\mbox{ } 

\ni The next stage involves dispensing with basing Physics upon some fixed topological space \cite{I89-Latt, I91}. 
A somewhat less radical version of that -- though it need not encapsulate stratified manifolds or lead to differentiable structure and the higher levels built upon that -- 
is dispensing with the idea of a fixed metric space \cite{I91}, and/or dispensing with some of the metric space axioms.  

\mbox{ } 

\ni What we do still had up to the previous level is the locally-Euclidean property, upon which rests the use of coordinate charts.
Some choices of individual coordinate therein might then be designated `time coordinate'.
Also it follows that there is still a fixed dimension; 3-$d$ is the smallest dimension that is sufficient to describe all experimentally/observationally confirmed Physics to date.  
Thus it is a good minimalistic choice that is not tied to making hypothesis.  

\mbox{ }

\ni However, it could well be that in a deeper theory dimension, orientability and whichever of openness, closedness and boundaries are emergent phenomena.  
These could indeed still be viewed as possibly background structure if treated as pre-determined inputs. 
The idea along these lines for now is concurrent inclusion of topologies which differ in these properties.
A problem with doing so is that transitions between them can involve singular spaces, and these are in general no longer manifolds.
Because of this, we consider the general topological space (or some restriction thereupon less stringent than manifoldness) 
to be the appropriate setting for this investigation, and thus difer it to the next Subsection.  

\mbox{ } 

\ni Are second-countability and Hausdorffness actually desirable properties for fundamental Theoretical Physics? 
Topological manifolds second-countability is a superset of operationally physical. 
This is because Physics in practise concerns only finite entities due to the nature of observations. 
Thus the second-countable property might be subjectible to some weakening.
Finally, topological manifolds are Hausdorff.
Hausdorffness is a great enabler of Analysis, so removing it induces technical difficulties.
Unfortunately there are are strong reasons to not want to always adhere to Hausdorffness in doing Physics. 
In particular, at the level of auxiliary spaces that feature in the Principles of Dynamics, 
it is relevant to note that neither local-Euclideaness nor Hausdorffness are in general inherited under quotienting. 
See Sec \ref{Spacetime-All} for an example. 
Finally, I leave for a future occasion consideration of whether Relationalism or Background Independence have any more precise say on 
where the fulcrum of Sec 4's balance between too large and too small, that is usually resolved by second-countability and Hausdorffness, is to lie.

\subsection{Background Independence at level of generally singular/stratified manifolds}\label{Strat-Man}

\ni This distinct specialization within topological spaces is the really desirable one as regards spaces of spaces at the more conventional upper levels of mathematical structure.  
Three strategies upon finding stratified manifolds are as follows. 

\mbox{ }  

\ni {\it Excise}, which is crude and unphysical. 

\mbox{ }  

\ni {\it Unfold} \cite{Fischer86}, but what is physical significance of the unfolding and is it unique. 

\mbox{ }  

\ni {\it Accept}, which requires harder mathematics (see below). 

\mbox{ }

\ni Each of topological-only stratification (Program A) and topological-and-(Riemannian-)metric level stratification (Program B) are interesting.

\subsection{Better patching as motivated by stratified manifolds}\label{Patch}

An evolving manifold in geometrodynamics is, at the topological level a bundle structure (upon excising symmetric configurations). 
Fibre bundles \cite{AMP} have the same fibre at each base point.
One can extend to attaching different objects at each base point by passing to the theory of {\bf sheaves} \cite{Sheaves}.
Loosely speaking, sheaves are tools for tracking locally defined entities by attachment to open sets within a topological space. 
They generalize fibre bundles by allowing for heterogeneity in the objects playing the role of fibres.
%
%

\mbox{ }

\ni Sheaves are then for instance well-suited for treatment of stratified configuration spaces and phase spaces \cite{BanaglBook}. 
%

\mbox{ }

\ni Fibre bundles' notion of global section generalizes to the case of sheaves. 
Thus sheaves present a way of formulating obstructions that generalizes the topological treatment via fibre bundles of a number of already understood obstructions.
In each case this points to an associated notion of cohomology.
In the case of sheaves, this is indeed termed a {\bf cohomology of sheaves} 
(this paper's use of this appears to be conceptually unrelated to the much earlier appearance of this notion in Penrose's twistor program \cite{Twistor}.

\mbox{ } 

\ni The other constituent part of a stratifold \cite{Kreck} is an algebra of continuous functions, which can be interpreted as an algebra of global sections in a sheaf-theoretic sense. 
%
%
Finally passing from manifolds to stratifolds simplifies singular bordisms to ordinary homology.  
%

\mbox{ } 

\ni {\it Topoi} (Sec \ref{Last}) are also a potentially useful tool as regards provision of superior patching methodology. 

\mbox{ }

\ni Thus, much as occurred with fibre bundles in the 1970's, a further generation of global/topological issues with physical theories could come to be understood in terms of 
sheaves and/or topoi.
See Sec \ref{Last} for a first quantum-level example of such a success.

\subsection{Background Independence and Problem of Time with even less structure}

A further idea here is {\it topologenesis}, in the distinct sense of topological spaces emerging from more general collections of subsets.

\mbox{ } 

\ni One can also choose to keep causal relations in the absense of manifoldness; the Causal Sets Approach \cite{Sor97} incorporates this using the mathematics of posets.  

\mbox{ } 

\ni As regards Spacetime Construction, in the Program B) version, something which becomes or approximately looks like a manifold in a limit.  
If classical spacetime is not assumed, one needs to get it back in a suitable limit. 
This can be hard; in particular, the less structure is assumed, the harder it is.
Examples of quite sparse structure are as follows.  

\mbox{ } 

\ni 1) Assuming just a discrete model of space. 

\ni 2) See e.g. \cite{RiWa} for recent Spacetime Construction advances in the Causal Sets Approach to Quantum Gravity.  

\mbox{ } 

\ni Here the spacetime causal order from GR and the path integral of QM are kept plus one new hypothesis: that spacetime is fundamentally discrete. 
The first and third of these assumptions are already held to apply at the classical level.  
The causal set replaces both the metric and the underlying topology all in one step.
That is relevant to the spacetime version of Topological Relationalism.  
In this approach, label invariance and growth order invariance take the place of coordinate invariance for causal sets.
In this approach also, time passing is an unceasing cascade of birth events.

\mbox{ } 

\ni Going one level further down, one might let the cardinality of the set that the topological spaces are based upon itself classically evolving and quantum-maechanically fluctuating.

\section{Notions of localization and information for Records Theory}

Example 1) Notion of distance criteria for locality in Records Theory apply at the level of Riemannian geometry.
Some such notions of distance were given by Kendall and Barbour for particle configurations in flat space \cite{Kendall, B94I, FileR}. 

\mbox{ }

\ni Example 1) This `metric' criterion descends to reduced configuration space.
The above comparers, alongside DeWitt's comparer \cite{DeWitt70} and a variety of other notions of distance between shapes were reviewed in \cite{FileR} for a wide range of theories.
This range includes 1) geometrodynamics, for which the comparers fail to be distances by indefiniteness of the configuration space metric. 
2) Conformogeometrodynamics, for which positive definiteness permits the interpretation of these comparers as distances to be regained.
The other techniques include `inhomogeneity comparers' \cite{FileR} (these can also be used between manifolds of different spatial topology). 
These are clearly also a type of shape comparer, just built out of diferent structures from the above three comparers.
A review of the structural elements involved in notions of distance between shapes can be found in \cite{FileR}.  

\mbox{ } 

\ni Example 3) This `metric' criterion also clearly descends to metric spaces and spaces of metric spaces (as per Sec 5).

\mbox{ } 

\ni Major Example 4) This criterion can furthermore then be replaced by topological notions of locality such as conditions for whether two neighbourhoods overlap 
as used e.g. in \v{C}ech cohomology. 

\mbox{ }

\ni {\bf \v{C}ech cohomology}.
The basic idea here is to model a topological space $\tau$ by open covers of it.
Doing this turns out to work well for `good covers':
\beq
\mbox{covers for which every open set and every finite intersection thereof are contractible }.
\label{Good-Cover}
\eeq
\ni One also defines the {\it nerve} ${\cal N}({\cal O})$ of each cover ${\cal O}$ as the simplicial complex built as follows.

\mbox{ }

\ni i) allotting 1 vertex per element of ${\cal O}$.

\ni ii) Allotting 1 edge per pair of open covers with non-empty intersection.

\ni iii) Continue through with this pattern to allotting 1 $k$-simplex per \{$k$ + 1\}-fold of open covers with non-empty total intersection
%

\mbox{ } 

\ni Then (\v{C}ech cohomology of $\tau$) = \big(simplicial cohomology of ${\cal N}({\cal O})$ for ${\cal O}$ of form (\ref{Good-Cover})\big).  
This is an example of a model space being a good model space by having properties of the underlying space that it is a model of.
In this case, the cohomological operation passing between open covers is refinement (or in the opposite sense coarsening) of open covers.

\mbox{ }

\ni Note that by this level, some aspects of Records Theory being a local pursuit require discussion. 
Many such set-ups can be envisaged as built out of local components.
This is the case for instance in attempt to match `multiple images' or `circles in the sky', since each is a localized part of the observational data.
It is also the case as regards the neighbourhoods that underlie the \v{C}ech cohomology example, 
which moreover are {\it locally compared} since the quantification is whether these overlap. 
{\sl Records Theory's locality criterion survives passage from a metric notion of locality to a topological spaces notion of locality.}  
Some parts of the `multiple images' and `circles in the sky' programs can be viewed as special cases of local Records Theory.
\v{C}ech cohomology can also be viewed in this light, as a mathematically well-defined class of problems that can be viewed as topologically local Records Theory.  

\mbox{ }

\ni Example 5) Furthermore, sheaf cohomology reduces to the \v{C}ech cohomology \cite{Cech} in the case of paracompact Hausdorff spaces \cite{Brylinski}, 
%
%
but the two are different on other and more general spaces. 
This points to a substantial further generalization from a \v{C}ech cohomology based Records Theory to a sheaf cohomology based Records Theory.  

\mbox{ } 

\ni Example 6) Finally, notions of locality can indeed also be defined on lattices, including thus the lattice of topologies on a fixed set.  
Since graining of topologies and graining of, more specifically, covers for topologies is tied to lattices, 
it no surprise that the space of all topologies on a fixed set is a lattice.   
Also, level by level treatment of graining in mathematics makes it clear that graining consequently occurs on configuration space (in timeless Records Theory),
 on phase space (standard Statistical Mechanics) and on spaces of histories.

\mbox{ } 

\ni As regards notions of information (some of which are up to a sign notions of entropy), 
see e.g. \cite{FileR} at the level of metric geometry and \cite{Top-Ent} at the metric space and topological space levels.

\mbox{ } 

\ni In outline for this more unfamiliar case of topological spaces, the notion here is based on covers.
Information consists of {\sl which subsets overlap}. 
{\sl Cover refinement} plays the role of fine graining.

\mbox{ } 

\ni Notions of entropy for sets have also been considered.

\section{Stochastic inputs into Records Theory}

\subsection{Metric geometry level}

First recollect Sec 3's shape statistics example that corresponds to relational mechanics.
Here to remain within metric manifolds rather than stratified manifolds, one considers simple special cases, such as the 1- and 2-$d$ cases.
For these a fully fledged SSSS is available.

\subsection{Topological manifold level}

Some simpler cases have been established in the Theoretical Physics literature for some time, namely those probing topology of the universe 
(meaning large-scale shape).\footnote{\ni By Fig \ref{Big-Superspace}.d-e), 
papers on the topology of the universe are really about a large scale approximate notion of topology that has not necessarily yet been quantified.
This is some new kind of mathematics: topology itself coarsened by length concepts, so that, somehow, large handles or tubes count, and small ones do not. 
`Large' here is with respect to the probing capacity of the observers.   
On some occasions, global effects will serve to discern which universe one is in, but I do not expect these to always be the case.}  
%
One idea here concerns recurring patterns.  

\mbox{ } 

\ni Example 7) Multiple images: if the universe is small enough we would see multiple copies of the same astrophysical objects, allowing for these images to correspond to different times. 

\ni Example 8) Circles in the sky, allowing for the universe to close up on scale bigger than Hubble radius and yet still imprint evidence of closing up within the cosmological horizon.

\mbox{ } 

\ni Major Example 4) Additionally, there has been more general recent work on in the Mathematics literature on stochastic topology by Niyogi, Smale and Weinberger \cite{NSW08}.
Some further useful tools are making use of random simplicial complexes, and(\v{C}ech co)homology.
This serves to assess the topology of an approximately given configuration.

\mbox{ } 

\ni Finally note that \v{C}ech cohomology transcends to the topological space level as well.    
On the other hand, applied to the present level of structure, \v{C}ech cohomology is well known to reduce to de Rham cohomology.

\subsection{Examples, including at the stratified manifold level}

\ni Finite stochastic geometry does range this far; the mechanics example of Sec 3 already establishes that stochastic geometry includes the finite stratified manifold arena.  

\mbox{ } 

\ni Considering Superspace($\bupSigma$) and CS($\bupSigma$) in this manner is subject to practical limitations.
Actually infinite-$d$ is not an issue in this case: stochastic treatment of Banach spaces and Fr\'{e}chet spaces is known \cite{P12}, 
including for Fr\'{e}chet spaces of the more specific type used to model Superspace($\bupSigma$).
However, I am not aware of this being extended to Banach or Fr\'{e}chet {\sl manifolds} much less to a suitable class of Fr\'{e}chet stratified manifolds.

\mbox{ }

\ni In homogeneous GR, alias minisuperspace, 
Diagonal minisuperspace has a flat shape space (space of anisotropies = `mini CS'), so it is too simple for nontrivial application of geometrical statistics.
Nondiagonal minisuperspace is, however, curved.

\mbox{ }

\ni Midisuperspace (inhomogeneous but somewhat symmetric GR) and/or 2 + 1 GR are more tractable than full 3 + 1 GR.  
The inhomogeneous perturbation model is furtherly simpler than the full 3 + 1 GR model, and can be seen as a replacement of the point particles of relational mechanics
by small inhomogeneous lumps within a GR framework. 

\mbox{ } 

\ni Tractability can also depend on the underlying manifold, as per e.g. the examples \cite{FM96} in which Superspace($\bupSigma$) is just an orbifold or even a manifold

\subsection{General Topological Space Level}\label{T}

\ni Example 5) As a further generalization, sheaf cohomology becomes distinct from \v{C}ech cohomology away from paracompact Hausdorff spaces.
This leads to a further range of records theories. 
Sheaf cohomology \cite{Iversen86} is widely computationally useful.

\mbox{ } 

\ni Example 8) (somewhat more restrictive). 
As regards working at the topological space level, Isham \cite{I84b} already pointed to Kendall's \cite{Kendall74} theory of `random T-sets'; 
this means he was making some contact with classical Probability Theory.
This work of Kendall's is within a carrier space ${\cal C}$ that is taken to be `H2LC'.
A {\it trapping system} ${\cal T}_{\sfT}$ for a random set $\mX \subset {\cal C}$ is then a collection of subsets with the following properties.   

\mbox{ }

\ni Trap-1) ${\cal T_{\sfT}} \neq \emptyset$, 

\ni Trap-2) $\bigcup_{\sfT} {\cal T}_{\sfT} = {\cal C}$, so the ${\cal T}_{\sfT}$ are a cover of ${\cal C}$. 

\ni Trap-3) To each ${\cal T}_{\sfT}$ one can associate a countable system ${\cal S}_{\sfS}({\cal T}_{\sfT})$ of subtraps (local countability).

\ni Trap-4) If $x \in {\cal T}_{\sfT}$, then $x$ belongs to a ${\cal C}$-trap whose ${\cal T}_{\sfT}$-closure is covered by ${\cal T}_{\sfT}$.   

\mbox{ } 

\ni The carrier space ${\cal C}$ being H2LC causes it to coincide with the type of space used in the more recent stratifold construct, which I have already argued to cover 
a number of physically relevant configuration spaces.\foo{Though as far as I know, 
neither Kendall's random sets nor stratifolds are vetted for use in infinite-$d$ cases based on in particular Fr\'{e}chet spaces, so this is not a saving grace as regards what follows.}
%
Finally, I indeed identify Kendall's theory of random sets as a further example of a classical Records Theory. 

\mbox{ }
	
\ni Example 6) Isham also considered the set of topological spaces on a fixed set $\mX$ to be a viable analogue of configuration space \cite{I89-Latt, I89-Rev}. 
Then as regards stochastic treatment of lattices, some well-known examples include random points on a square lattice \cite{Roach} and Percolation Theory \cite{BRBook}.
{\sl Stochastic treatment of a more suitably general notion of a lattice has been developed by Molchanov} \cite{Molchanov}.
This is a step in the right direction as regards considering a stochastic treatment of the lattice of topologies on a fixed set itself.
In cases considering correlations within a lattice, Molchanov's approach can be viewed as a Records Theory. 
If this methodology can be applied to the lattice of topologies on a fixed set, that case of it would furthermore be a type of shape structure (generalizer of shape geometry) 
and shape statistics, for a topological space level notion of shape.

\mbox{ } 

\ni Note that Examples 5) and 9) are space by space considerations, whereas Example 6) is on the whole space of spaces.

\subsection{Level of sets}

This is straightforward due to the lack of additional structures.

\subsection{Level of collections of subsets}

This is interesting, though for now I am not aware of work in this direction.
E.g. how probable is it for a random collection to be a topology? 
A cover? 
A good cover? 
%
%
A $\sigma$-algebra? 
A trapping system?

\section{Quantum Background Independence and the Problem of Time}\label{QBI}

Sec 2 serves well enough for this at the metric and differentiable manifold geometry levels.  
What happens then at the deeper levels?

\mbox{ }

\ni I begin by recollecting some useful tricks.    

\mbox{ } 

\ni 1) If a system's configuration space is of the form 
\be
\FrQ \,\, = \, \, \FrG/\FrH \mbox{ } , 
\ee
i.e. a homogeneous space for $\FrG$ a group and $\FrH$ a subgroup, 
then variouCs simplifications occur and the powerful induced representation techniques due to Mackey become available \cite{I84}.
This covers e.g. textbook QM, various simple quantum models in \cite{I84}, relational mechanics models \cite{FileR}, the Poincar\'{e} group of standard QFT 
and, to some extent, the diffeomorphisms \cite{I84}. 
Moreover, this trick can indeed be extended \cite{I03} to the case in which $\FrQ$ is a {\it generalized} configuration. 

\mbox{ } 

\ni 2) Some examples possess enough similarities to the Fock space approach familiar from QFT.

\subsection{General issues}\label{QM-Gen}

1) $\scC\scH\scR\scO\scN\scO\scS$       gives rise to a frozen $\widehat{\scC\scH\scR\scO\scN\scO\scS}\Psi = 0$, supplemented by 
2) $\scS\scH\scU\scF\scF\scL\scE_{\sg}$ giving rise to $\widehat{\scS\scH\scU\scF\scF\scL\scE}_{\sg}\Psi = 0$
3) Quantum level Constraint Closure is demanded, noting that classical closure does not guarantee this and (11)'s functional evolution problem.
4) Quantum beables are sought for.
5) Attempted use of spacetime-type notions now has extra path integral connotations, including $Aut(\FrG_{\sS})$-independent measures. 
6) Quantum level Foliation Independence and 7) Spacetime Construction are not assured by derived mathematical results even at the usual levels of structure.
At any level of structure, it remains desirable for timefunctions and all other steps involved in the above exposition to be 8) {\bf Globally Valid}.
%
%
Finally 9) {\bf Reconcileability of Multiplicity} would apply in cases in which a multiplicity of elsewise-valid timefunctions occur.

\subsection{Topological manifold relationalism at the quantum level}\label{QM-Top}

In Wheeler's spacetime foam picture \cite{WheelerGRT}, quantum fluctuations could be expected to cause the metric to change signature and alter the topology of space.
Thus Quantum Theory might tolerate, motivate even, inclusion of more general metrics such as degenerate or singular ones amongst the configurations of one's theory. 
On these grounds, the classical-level inclusion of degenerate beins in Ashtekar variables formulations on other grounds is perhaps acceptable,  
and similarly the consideration of plain rather than affine geometrodynamics at the quantum level.  

\mbox{ }

\ni Wheeler's considerations came from the idea of applying Feynman path integrals to Quantum Gravity. 
Perhaps then now sum over histories entails sums over topologies, and transition amplitudes for spatial topology change could then be computed.
There are then issues about more specifically which topologies to include, and which metrics thereupon.
On the other hand, topology change is more problematic to study from a canonical perspective.

\mbox{ }

\ni QM backs the extremal action's contribution value, albeit only from this contributing more rather than from it being sole contributor.  
This is a discrete parameter extremization of actions selecting dynamical paths involves extremizing continuous parameters.  
At the very least one would have to use an extended version of the calculus of variations to justify this, and how to do so is not clear. 
The biggest action provides the most probability, 
but other case of sizeable action contribute too and could dominate eg though being more numerous: quantum smoothening out of classical extremum quantities. 

\mbox{ }

\ni If pinching off is allowed, a quantum-level microscopic problem ensues to virtual pairs pinching all over the place.  

\mbox{ }

\ni Isham \cite{I89-Rev} has also pointed out that functional integrals involve distributions 
-- quantum theory `roughens up' spaces so why not sum over differentiable manifolds alongside manifolds with singularities in place of just over differentiable manifolds?
Also, Isham did not just view Fig \ref{Bigger-Set-2}'s levels of structure as mathematics underlying physics, but furthermore as a sequence of structures to quantize in turn.  
Already in 1989, he wrote \cite{I89-Rev}
``{\it Nonwithstanding the current popularity of differential geometry, my strong belief is that its days are numbered, at least so far as the subject 
of Quantum Gravity is concerned.
Smooth manifolds and local differential equations belong primarily to the world of classical physics and I do not believe that these are appropriate tools with which to probe the 
structure of spacetime (in so far as this is a meaningful concept at all) near the Planck length.
At best, they are likely to be applicable in the semiclassical limit of the quantum theory of gravity (whatever that may be) and a lot more thought needs to be given to 
the question of which mathematical structures are really relevant for discussing the concepts of space and/or time in the ``deep" quantum region.}".
This echoes \cite{BI99} Riemann's much earlier quotation {\it ``Now it seems that the empirical notions on which the metrical determinations of space are founded, 
the notion of a solid body and of a ray of light, cease to be valid for the infinitely small. 
We are therefore quite at liberty to suppose that the metric relations of space in the infinitely small do not conform to the hypotheses of geometry; 
and we ought in fact to suppose it, if we can there by obtain a simpler explanation of phenomena."} 

\mbox{ } 

\ni One generalization of QFT (often via CFT) is TFT.
Consider especially the case of Chern--Simons Theory \cite{TFT}.  
This is metric-free.
This and topology change in GR are both connected to Morse Theory.   
This reflects Morse Theory's affording a suitable simple model of mildly singular spaces.
%
%
There attestedly is no problem with having Principles of Dynamics actions at this level.  

\mbox{ } 

\ni As an example of some progress with kinematical-level detail, Gibbons and Hawking \cite{GH92} derived a selection rule for the path integral approach to Quantum Gravity.
Namely that handle creation and annihilation must involve pairs of handles.

\subsection{Quantum-level Metric Space and Topological Space Relationalism and beyond}

\ni In his quotation in the previous section, Isham is in part concerned that point-set theory is used even when points are held to be physically meaningless.
E.g. why should probabilities belong to the real interval [0, 1]? 

\mbox{ }

\ni At least in their current forms, most of the currently pursued approaches to Quantum Gravity assume continuum notions at some level or another \cite{BI99}.  
However, beyond a certain point, 
use of continuum notions (on which manifolds are based: spacetime, space, Principles of Dynamics spaces, Lie groups, probabilities...) becomes a background presumption.
So are the choice of function spaces (most notably standard Hilbert spaces at the quantum level) 
and the the standardly-adopted suppositions that one's mathematics can be rooted in set theory and that standard `yes/no answer' binary logic is in use.  
Thus discrete approaches can be viewed as a challenge to manifolds, 
and such as lowering standard differentiability or passing to use of Sobolev spaces as a lesser challenge to part of the assumptions at the level of differentiable manifolds.
Metric and topological spaces include discrete as well as continuum cases, and Isham investigated quantization at each of these levels in \cite{I91}.

\mbox{ } 

\ni In fact, the suggestion of quantizing distance itself \cite{IKR} goes back to Wheeler \cite{WheelerGRT}.  
Quantizing distance requires the affine insight for the same reasons as usual.
One then quantizes with the space of norms playing the role of configuration space (c.f. how Isham and Linden \cite{IL} also treated histories as configurations).  

\mbox{ }

\ni One continues to have \cite{I89-Wig} s semidirect product of groups in this case, permitting once again the use of Mackey Theory to extract representations.
One has a square root to contend with, with the entity inside being concentrated on a curve, so Isham is concerned that this is singular enough for the root to cause difficulties.  
Moreover, the above two examples turns out to have enough parallels for QFT to permit a Fock space based approach.  
In particular, analogues of creation and annihilation operators can be defined for these. 

\mbox{ } 

\ni Finally in \cite{I89-Latt} Isham considers various attitudes to time at the topological level, including timelessness, discrete time steps, and path integrals.   
On the other hand, in \cite{I89-Rev}, Isham viewed time as a continuous label, and along the lines of an internal time, at least in the semiclassical limit.
None of theese correspond to the emergent Machian time or records approaches that are exposited in the current Article.

\subsection{Quantization at yet deeper levels of structure}

\ni A first example of this is quantization of causal sets.       
There are some treatments of this by Sorkin \cite{Sor83, Sor91} and more recently by Isham also \cite{I03}.  

\mbox{ } 

\ni As a further question at the level of sets themselves, should the cardinality of the underlying set itself be subjected to quantum fluctuations?
Isham \cite{I03} also considers a case of quantization at the level of sets themselves. 

\mbox{ } 

\ni Note however that these examples of Isham's lie within the paradigm of Sec \ref{Last}, details of which lie outside of the scope of the present article.

\subsection{Situations in which deeper levels of structure can be neglected}

The semiclassical Quantum Cosmology regime does not involve fluctuating topologies or worse, protecting us from this tower for some practical purposes. 
Of course, the most interesting questions in Quantum Gravity are however about more full regimes than that.

\section{Further questions}

\subsection{Concerning Records and Histories}

\ni Question 1) Given the classical advances involving stochastic mathematics, what of theories of quantum records in such terms? 

\mbox{ }

\ni Question 2) What form does a classical histories theory take at each level? 

\mbox{ }

\ni Question 3) What of a theory of quantum histories at each level?

\mbox{ }

\ni Question 4) What of a combined emergent time, histories and records seheme at each level?

\mbox{ }

\ni Question 5) Indeed, can one justify the usual use of `configurations' to hold truer to some mark than the extended use?

\subsection{Concerning further quantum-level possibilities offered by categories and topoi}\label{Last}
%

Category Theory and Topos Theory offer alternatives to the `equipped sets' paradigm of the levels of mathematical structure that is conventionally used in Theoretical Physics.
As regards quantization, Isham \cite{I03} carried this out on {\it small categories}.\footnote{This is a substantial limitation on applications, 
since most of the categories of interest as regards the mathematics underlying Theoretical Physics are at least locally small.}
%
Here the objects are generalized configurations and the morphisms alias arrows as generalized momenta.
The $\FrQ \,\, = \, \, \FrG/\FrH$ trick is again employed in these examples, along with some uses of creation--annihilation operator analogues.

\mbox{ } 

\ni Question 6) Carry out a relational and background-independent analysis of this approach, as well as an analysis of the corresponding classical dynamics and SSSS.

\mbox{ } 

%
\ni {\bf Topoi}.
One could base one's physical theories on topoi instead of sets {\sl and} with multi-valued and contextual logic.  
Here answers can be multi-valued rather than YES/NO, as well as this {\it valuation} `differing from place to place'.
For these are geometrical logics whose valuations are now along the lines of locally-valid charts in differential geometry to Boolean logic's globally valid valuations being somewhat like 
flat space.  
Another of the many ways of envisaging a topos is as a category with three extra structures that give it some properties similar to those of sets. 
These are \cite{ToposRev, Lawvere} 1) finite limits and colimits 2) power objects and 3) a subobject classifier 
(which in the case of sets alongside the standard logic is the means of allotting yes or no answers to propositions).  
%
%
By being more set-like, topoi may be more suitable than categories as regards replacing sets as regards primality.  
See in particular \cite{Johnstone} for further perspectives on what topoi are. 
Indeed, they are notorious for being a coincidence of multiple perspectives, in a more mathematically rigorous parallel to how Wheeler argued there to be `many routes to GR'.
That is potentially of interest in seeking a quantum version of GR, though two theories being rich in the admission of multiple perspectives does not suffice for them to be related.

%
\ni Topos approaches are likely to be useful in Quantum Theory or some replacement or generalization thereof, whether or not this ameliorates the clash with gravitation.
Isham considered using topoi to upgrade quantization in \cite{I05, ToposRev, I10} 
(some co-authored with Doering; see \cite{Landsman} for Heunen, Landsman and Spitters' alternative approach).
N.B. this use of Topos is more subtle than 'quantization on'. 
Rather, conceptual issues in (interpretation of) QM can be cast in useful new geometrical light using Topos Theory.

\mbox{ }
   
\ni As per Sec \ref{Patch}, sheaves are the basis for more general patching constructs.   
{\bf Presheaves} are a possible less structured alternative to sheaves. 
The most common set-valued presheaves on a category $\textgoth{C}$ are functors $\textgoth{f}: \textgoth{C}^{\so\sp} \rightarrow$ \mbox{\textgoth Sets}, 
often written as $\hat{\textgoth{C}} = \mbox{\textgoth Sets}^{\mbox{\scriptsize\textgoth{C}}^{\to\tp}}$.  
Here `op' denotes oppositeness, meaning that the functor involved maps in the opposite direction (the two directions are termed covariant and contravariant respectively). 
Presheaves become sheaves if a technical condition concerning locality and another guaranteeing that compatible sections can be glued together can be appended. 
Thus sheaves carry global patching connotations beyond those of presheaves.  

\mbox{ } 

\ni QM ends up involving the topos $\mbox{\textgoth Sets}\mbox{}^{\mbox{\scriptsize\textgoth V}(\sH\si\sll\sb)^{\to\tp}}$. 
I.e. the topos of contravariant functors on the poset category \textgoth{V}(Hilb) of commutative subalgebras of the algebra of bounded operators on the Hilbert space of the system, Hilb. 
Presheaves and sheaves are equivalent in this case's working.
This approach succeeded in reformulating the Kochen--Specker Theorem, schematically \cite{IBook}
\beq
\mbox{function(valuation($\widehat{A}$)) \mbox{ } need not be the same as \mbox{ } valuation(function($\widehat{A}$)) for Hilbert spaces of dimension $> 2$ } ,    
\eeq
in terms of a presheaf on the category of self-adjoint operators having no global elements. 
This presheaf presentation suggests a generalization of the definition of valuation that is more suitable to Quantum Theory.

\mbox{ }

\ni A connection has also already been made between Histories Theory \cite{IL} and the Topos approach \cite{I97}, 
and another has been sketched between Timeless Records Theory and the Topos approach (\cite{FileR}, based on interpreting \cite{ToposRev} in this manner).   

\mbox{ } 

%
\ni One longstanding, if nebulous, idea in Quantum Gravity has been termed `field marshal covariance' \cite{Strominger} 
(as in outranking general covariance by its having an even wider scope). 
In this regard, I mention that Heunen, Landsman and Spitters have formulated a notion of {\it general tovariance} \cite{Tovariance} (a topos counterpart of general covariance), 
which might in this way be a realization of `field marshal covariance'. 
[Non-commutative geometry's version of general covariance is also more general than GR's, as is supergravity's.
However Topos Theory greatly further outstrips these theories as regards mathematical generality and non-assumption of `standard' mathematical structures in physics that are in fact 
only standard due to being rooted in absolutist and elsewise background-dependent assumptions.]

\mbox{ } 

\ni Question 7) Another possible future direction concerns the  main part of the Multiple-Choice problem -- the Gronewold--van Hove phenomenon. 
This is a type of obstruction, albeit a hitherto not well understood one.
Does recasting this in the language of sheaves or topoi clarify this phenomenon and whether it is ubiquitous (e.g. in some sense generic)?  

\mbox{ }

\ni Synthetic mathematics offers further alternatives \cite{Synth}. 
This is a top-down approach to mathematics from granting that one in fact has rigorous infinitesimals ab initio, one does not need many other structures.  
Note that top-downness is a more radical alternative to the `equipped sets' paradigm than in Sec \ref{Commentary}, 
whose alternatives were merely based upon changing which entities are accorded the primary status prior to the onset of equipping them with further structures.
[Topos theory can also be regarded as top-down in some ways, eg in allotting primary significance to regions rather than to points which regions are subsequently taken to be composed of.] 

\mbox{ } 

\ni Question 8) Can Relationalism or Background Independence more generally discern between standard and synthetic mathematics being used at the deeper levels of Theoretical Physics?

\mbox{ } 

\ni Question 9) Similarly, might Relationalism or Background Independence be applicable as criteria to distinguish between standard `equipped sets' mathematics and categorical or 
topos-theoretic mathematics being used to model the deeper levels of Theoretical Physics?

\mbox{ } 

\ni For instance, might it in fact be standard mathematics' notions such as points having primary ontological status, or the notion of open sets, that are artificial?

\mbox{ }

\ni {\bf Acknowledgements}: 
E.A. thanks close people. 
Chris Isham for discussions.  
Jeremy Butterfield, John Barrow, Marc Lachi$\grave{\me}$ze--Rey, Malcolm MacCallum, Don Page, Reza Tavakol and Paulo Vargas-Moniz for help with my career.

\vspace{10in}


\end{document}